\documentclass[utf8]{frontiersFPHY}
\usepackage[english]{babel}
\usepackage[T1]{fontenc}
\usepackage[colorlinks=true,linkcolor=black,citecolor=black,urlcolor=black]{hyperref}
\usepackage{bm,bbm,amssymb,amsmath}
\usepackage{graphicx}
\usepackage{multirow,booktabs,dcolumn}
\usepackage{isotope}
\usepackage{physics}
\usepackage[capitalize]{cleveref}
\usepackage{nicefrac}
\usepackage{ulem}
\usepackage[onehalfspacing]{setspace}
\setcitestyle{square}

\def\keyFont{\fontsize{8}{11}\helveticabold }
\def\firstAuthorLast{Gandolfi \textit{et~al.}}
\def\Authors{Stefano Gandolfi\,$^{1,*}$, Diego Lonardoni\,$^{1,2}$, Alessandro Lovato\,$^{3,4}$, and Maria Piarulli\,$^{5}$}
\def\Address{$^{1}$Theoretical Division, Los Alamos National Laboratory, Los Alamos, NM 87545, USA \\
$^{2}$Facility for Rare Isotope Beams, Michigan State University, East Lansing, MI 48824, USA\\
$^{3}$Physics Division, Argonne National Laboratory, Argonne, IL 60439, USA\\
$^{4}$INFN-TIFPA Trento Institute of Fundamental Physics and Applications, 38123 Trento, Italy\\
$^{5}$Physics Department, Washington University, St Louis, MO 63130, USA}
\def\corrAuthor{Stefano Gandolfi, Theoretical Division, Los Alamos National Laboratory, Los Alamos, NM 87545, USA}
\def\corrEmail{stefano@lanl.gov}

\begin{document}
\onecolumn
\firstpage{1}

\title[Nuclei: QMC and $\chi$EFT Interactions]{Atomic nuclei from quantum Monte Carlo calculations with chiral EFT interactions} 

\author[\firstAuthorLast ]{\Authors}
\address{}
\correspondance{}
\extraAuth{}

\maketitle

\begin{abstract}
Quantum Monte Carlo methods are powerful numerical tools to accurately solve the Schr\"odinger equation
for nuclear systems, a necessary step to describe the structure and reactions of nuclei and nucleonic matter 
starting from realistic interactions and currents. These \textit{ab-initio} methods
have been used to accurately compute properties of light nuclei -- including their spectra, moments, and transitions -- and the equation of state of neutron and nuclear matter. In this work we review selected results obtained by combining quantum Monte Carlo methods and recent Hamiltonians constructed within chiral effective field theory.

\tiny
\keyFont{ \section{Keywords:} Quantum Monte Carlo methods, variational Monte Carlo, Green's function Monte Carlo, auxiliary field diffusion Monte Carlo, chiral effective field theory, nuclear Hamiltonians, nuclear structure} 
\end{abstract}

\section{Introduction}
The study of nuclear properties as they emerge from the individual interactions among protons and neutrons is a fascinating long-standing problem, subject of both theoretically and experimentally research activities. From a theoretical point of view, a truly \textit{ab-initio} description of nuclei is still very challenging at present. The underlying theory of strong interactions, Quantum Chromodynamics (QCD), that describes how quarks and gluons interact to form nucleons and nuclei, in the low-energy regime is non-perturbative in its coupling constant. Despite remarkable progresses~\cite{Winter:2017bfs,Chang:2017eiq}, realistic computations of many-body nuclear systems in terms of the fundamental degrees of freedom of QCD -- quarks and gluons -- are still extremely challenging.

A more feasible approach to the problem consists in assuming that at the energy regime relevant to the description of atomic nuclei, quarks, and gluons are confined within hadrons. The latter are the active degrees of freedom at soft scales, and they interact among themselves through non-relativistic effective potentials that are consistent with the symmetries of QCD. The solution of nuclear many-body problems requires two main ingredients: an Hamiltonian that accurately models the interactions among the nucleons, and reliable numerical many-body methods to solve the corresponding Schr\"odinger equation.

Microscopic nuclear Hamiltonians, capable of reproducing 
nucleon-nucleon scattering data and the properties of few-body systems, have been successfully used to describe light nuclei. For example, the
highly-realistic Argonne $v_{18}$ two-body potential~\cite{Wiringa:1994wb} combined with the phenomenological Illinois-7 three-body force have been employed to predict several properties of nuclei up to $A=12$ with great accuracy~\cite{Carlson:2015}.
Several calculations of energies, rms radii, transitions, and densities turn
out to be in excellent agreement with experimental data.
The main limitation of these phenomenological Hamiltonians is that it is not clear how they can be systematically improved, and how to quantify theoretical, i.e., systematic, uncertainties related to the specific interaction model. Another approach that became very popular in the last two decades consist in deriving nuclear interactions within the framework of chiral Effective Field Theory ($\chi$EFT). The advantage of this approach is that it provides the necessary tools to systematically improve the interaction models, to estimate uncertainties related to the truncation of the chiral expansion, and to consistently derive electroweak currents.

Several many-body methods have been developed to numerical solve 
the many-body Schr\"odinger equation. Most of them rely on basis expansions, for example the coupled cluster method~\cite{Hagen:2014,Hagen:2014b},
the no core shell model~\cite{Barrett:2013}, the similarity renormalization group~\cite{Bogner:2010}, and the self consistent Green's function~\cite{Dickhoff:2004}. Each of these methods has distinct advantages, and many are able to treat a wider variety of nuclear interaction models. These many-body techniques are very effective and achieve  a good convergence only when relatively soft potentials are used.

Quantum Monte Carlo (QMC) methods are ideally suited to study strongly correlated many-body systems, and have no difficulties in treating ``stiff'' nuclear interactions, but are limited to nearly local nuclear potentials. For this reason, until fairly recently, the applicability of QMC methods was limited to phenomenological interactions, as $\chi$EFT Hamiltonians were typically written in momentum space. Over the past few years, the situation has drastically changed with the development of local $\chi$EFT potentials, both with~\cite{Piarulli:2014bda,Piarulli:2017dwd} and without explicit delta degrees of freedom~\cite{Gezerlis:2013ipa,Tews:2015ufa}, that have provided a way to combine an EFT-based description of nuclear dynamics with precise QMC techniques. In this work we will review selected results of nuclei obtained using QMC methods and chiral Hamiltonians.

\section{Nuclear interactions}
The microscopic model of nuclear theory assumes that nuclear systems can be described as point-like nucleons, whose dynamics is characterized by a non-relativistic Hamiltonian
\begin{align}
H=\sum_i T_i+\sum_{i<j}v_{ij}+\sum_{i<j<k}V_{ijk}+ \cdots\, ,
\end{align}
where $T_i$ is the one-body kinetic energy operator, $v_{ij}$ is the nucleon-nucleon ($N\!N$) interaction between particles $i$ and $j$, $V_{ijk}$ is the three-nucleon ($3N$) interaction between particles $i$, $j$, and $k$, and the ellipsis indicate interactions involving more than three particles. There are indications that four-nucleon interactions may contribute at the level of only $\sim100\,\rm keV$ in \isotope[4]{He}~\cite{Rozpedzik:2006yi} or pure neutron matter~\cite{Kruger:2013kua}, and therefore are negligible compared to $N\!N$ and $3N$ components. Hence, current formulations of the microscopic model do not typically include them (see, for example, reference~\cite{Carlson:2015}). 

The $N\!N$ interaction term in the nuclear Hamiltonian is the most studied of all, with thousands of experimental data points at laboratory energies ($E_{\rm lab}$) from essentially zero to hundreds of MeV. It consists of a long-range component, for inter-nucleon separation $r \gtrsim 2$ fm, due to one-pion exchange (OPE)~\cite{Yukawa:1935xg}, and intermediate- and short-range components, for, respectively, $1 \,{\rm fm} \lesssim r \lesssim 2\,\rm fm$ and $r \lesssim 1 \, \rm fm$, derived, up to the mid 1990's, almost exclusively from meson-exchange phenomenology~\cite{Stoks:1994wp,Wiringa:1994wb,Machleidt:2000ge}. These models fit the large amount of empirical information about $N\!N$ scattering data contained in the Nijmegen database~\cite{Stoks:1993tb}, available at the time, with a $\chi^2/{\rm datum}\simeq 1$ for $E_{\rm lab}$ up to pion-production threshold. Two well-known and still widely used examples in this class of $N\!N$ interactions are the CD-Bonn~\cite{Machleidt:2000ge} and the Argonne $v_{18}$ (AV18)~\cite{Wiringa:1994wb} potentials. 

The AV18 interaction is a local, configuration-space $N\!N$ potential that has been extensively and successfully used in a number of QMC calculations. It is expressed as a sum of electromagnetic and OPE terms and phenomenological intermediate- and short-range parts:
\begin{align}
v_{ij} =v^{\gamma}_{ij}+ v^{\pi}_{ij}+ v^{I}_{ij}+v^{S}_{ij}\, .
\label{eq:pot}
\end{align}
The electromagnetic term $v^{\gamma}_{ij}$ has one- and two-photon exchange Coulomb interaction, vacuum polarization, Darwin-Foldy, and magnetic moment terms, with appropriate form factors that keep terms finite at $r=0$ (see reference~\cite{Wiringa:1994wb} for more details). The OPE part includes the charge-dependent (CD) terms due to the difference in neutral ($m_{\pi_0}$) and charged pion ($m_{\pi_\pm}$) masses, and in coordinate-space it reads
\begin{align}
\label{eq:vpi_r}
v^{\pi}_{ij}=\left[ v^{\pi}_{\sigma \tau}(r) \, {\bm \sigma}_i \cdot {\bm \sigma}_j +
v^{\pi}_{t \tau}(r) \, S_{ij}\right] \,  {\bm \tau}_i \cdot {\bm \tau}_j+\left[ v^{\pi}_{\sigma T}(r) \, {\bm \sigma}_i \cdot {\bm \sigma}_j +
v^{\pi}_{t T}(r) \, S_{ij}\right]\,T_{ij}\ ,
\end{align}
where ${\bm \sigma}$ adn ${\bm \tau}$ are the Pauli matrices that operate over the spin and isospin of particles, and $S_{ij}= 3\, {\bm \sigma}_i \cdot \hat{\textbf r}_{ij} \,\, {\bm \sigma}_j \cdot \hat{\textbf r}_{ij}- {\bm \sigma}_i \cdot {\bm \sigma}_j$ and $T_{ij}=3\,\tau_{iz}\tau_{jz}-{\bm \tau}_i\cdot {\bm \tau}_j$ are the tensor and isotensor operators, respectively. The functions, $v^{\pi}_{\sigma \tau}(r)$, $v^{\pi}_{t \tau}(r)$, $v^{\pi,\rm}_{\sigma T}(r)$, and $v^{\pi}_{t T}(r)$ are defined as 
\begin{align}
\label{eq:OPEfuncs}
v^{\pi}_{\sigma\tau}(r)&=\frac{Y_0(r)+2\, Y_+(r)}{3}\,, &&v^{\pi}_{t\tau}(r)=\frac{T_0(r)+2\, T_+(r)}{3} \,, \nonumber \\
v^{\pi}_{\sigma T}(r)&=\frac{Y_0(q)- Y_+(r)}{3}\,, &&v^{\pi}_{tT}(r)=\frac{T_0(r)-T_+(r)}{3} \,,
\end{align}
where $Y_\alpha(r)$ and $T_\alpha(r)$ are the Yukawa and tensor functions given by
\begin{align}
Y_\alpha(r)= \frac{g_A^2}{12\,\pi}\,\frac{m^3_{\pi_\alpha}}{(2\,f_{\pi})^2} \,\frac{e^{-x_\alpha}}{x_\alpha}\,, \qquad
T_\alpha(r)=Y_\alpha(r)\left( 1+\frac{3}{x_\alpha}+\frac{3}{x^2_\alpha}\right)\,,
\end{align}
with $x_\alpha=m_{\pi_\alpha} r$, and $g_A=1.267$, $f_\pi=92.4\,\rm MeV$ being the axial-vector coupling constant of the nucleon and the pion decay constant, respectively. 

The intermediate-range region, $v^{I}_{ij}$, is parametrized in terms of two-pion exchange (TPE), based on, but not consistently derived from, a field-theory analysis of box diagrams with intermediate nucleons and $\Delta$ isobars~\cite{Smith:1976zc}. The short-range region, $v^{S}_{ij}$, is instead represented by spin-isospin and momentum-dependent operators multiplied by Woods-Saxon radial functions~\cite{Wiringa:1994wb}. 

The AV18 model can be written as an overall sum of eighteen operators $\left(\mathcal N=18\right)$
\begin{align}
v_{ij} = \sum_{p=1}^{\mathcal N}v^p(r_{ij})\mathcal O^p_{ij}\, ,
\label{eq:NN_pot}
\end{align}
where the first eight are given by
\begin{align}
\mathcal O^{p=1-8}_{ij}= \big[1, {\bm \sigma}_i \cdot {\bm \sigma}_j,S_{ij},\mathbf{L}\cdot\mathbf{S}\big]\otimes\big[1,{\bm \tau}_i \cdot {\bm \tau}_j \big] \, ,
\label{eq:oper}
\end{align}
with the spin-orbit contribution expressed in terms of the relative angular momentum $\mathbf{L}=\frac{1}{2 i} (\mathbf{r}_i -\mathbf{r}_j) \times (\bm\nabla_i -\bm\nabla_j)$ and the total spin $\mathbf{S}=\frac{1}{2}({\bm \sigma}_i+{\bm \sigma}_j)$ of the pair. There are six additional charge-independent operators corresponding to $p = 9 - 14$ that are quadratic in $L$
\begin{align}
\mathcal O^{p=9-14}_{ij}= \big[\mathbf{L}^2, \mathbf{L}^2\,{\bm \sigma}_i \cdot {\bm \sigma}_j,(\mathbf{L}\cdot\mathbf{S})^2\big]\otimes\big[1,{\bm \tau}_i \cdot {\bm \tau}_j\big] \, ,
\label{eq:operl2}
\end{align}
while the $p = 15 - 18$ are charge-independence breaking terms
\begin{align}
\mathcal O^{p=15-18}_{ij}=\big[T_{ij}, T_{ij}\,{\bm \sigma}_i \cdot {\bm \sigma}_j,T_{ij}\,S_{ij},\tau_{i,z}+\tau_{j,z}\big] \, .
\label{eq:opercb}
\end{align}

The AV18 model has a total of 42 independent parameters. A simplex routine~\cite{Nelder:1965zz} was used to make an initial fit to the
phase shifts of the Nijmegen partial-wave analysis (PWA)~\cite{Stoks:1993tb}, followed by a final fit direct to the database, which contains 1787 $pp$ and 2514 $np$ observables for $E_{\rm lab}$ up to $350\,\rm MeV$. The $nn$ scattering length and deuteron binding energy were also fit. The final $\chi^2/{\rm datum}=1.1$~\cite{Wiringa:1994wb}. While the fit was made up to $350\,\rm MeV$, the phase shifts are qualitatively good up to much larger energies, $E\leq 600\,\rm MeV$~\cite{Gandolfi:2013baa}. 

Simplified versions of these interactions, including only a subset of the operators in Equation~(\ref{eq:oper}), are available. For instance, the Argonne $v_8^{\prime}$ (AV8$^{\prime}$)
contains a charge-independent eight-operator projection, $\mathcal O^{p=1-8}_{ij}= \big[1, {\bm \sigma}_i \cdot {\bm \sigma}_j,S_{ij},\mathbf{L}\cdot\mathbf{S}\big]\otimes \big[1,{\bm \tau}_i \cdot {\bm \tau}_j \big]$, of the full $N\!N$ potential, constructed to preserve the potential in all $S$ and $P$ waves as well as the $^3D_1$ and its coupling to the $^3S_1$, while over-binding the deuteron by $18\,\rm keV$ due to the omission of electromagnetic terms~\cite{wiringa02}. The main missing features of these simplified interactions is the lack of terms describing charge and isospin symmetry breaking, as well as a slightly poorer description of nucleon-nucleon scattering data in higher partial waves. However, these contributions are very small, as outlined in reference~\cite{wiringa02}.

Already in the 1980s, accurate three-body calculations showed that contemporary $N\!N$ interactions did not provide enough binding for the three-body nuclei, \isotope[3]{H} and \isotope[3]{He}~\cite{Friar:1984ic}. In the late 1990s and early 2000s this realization was also extended to the spectra (ground and low-lying excited states) of light $p$-shell nuclei, for instance, in calculations based on QMC methods~\cite{Pudliner:1997ck} and in no-core shell-model studies~\cite{Navratil:2000gs}. Consequently, the microscopic model with only $N\!N$ interactions fit to scattering data, without the inclusion of a $3N$ interaction, is no longer considered realistic.

In addition to $N\!N$ forces, sophisticated phenomenological $3N$ interactions have been then developed. They are generally expressed as a sum of a TPE $P$-wave term, a TPE $S$-wave contribution, a three-pion-exchange contribution, and a $3N$ contact~\cite{Carlson:2015}.
More specifically, two families of $3N$ interactions were obtained in combination with the AV18 potential: the Urbana IX (UIX)~\cite{Carlson:1983kq} and Illinois 7 (IL7)~\cite{Pieper:2001ap} models. The UIX potential contains two parameters fit to reproduce the ground-state energy of \isotope[3]{H} and the saturation-point of symmetric nuclear matter, while the IL7 potential involves five parameters constrained on the low-lying spectra of nuclei in the mass range $A=3-10$. 

Despite their success in predicting a wide range of nuclear properties~\cite{Carlson:2015}, the phenomenological potentials suffer from several drawbacks. For example, the resulting AV18+IL7 Hamiltonian leads to predictions of $\approx 100$ ground- and excited-state energies up to $A=12$ in good agreement with the corresponding empirical values. However, when used to compute the neutron-star equation of state, such Hamiltonian does not provide sufficient repulsion to guarantee the stability of the observed stars against gravitational collapse~\cite{Maris:2013rgq}.
On the other end, the AV18+UIX model, while providing a reasonable description of $s$-shell nuclei and nuclear matter properties, it somewhat underbinds light $p$-shell nuclei.

Thus, in the context of the phenomenological nuclear interactions, we do not have a Hamiltonian that can explain the properties of all nuclear systems, from $N\!N$ scattering to dense nuclear and neutron matter. Furthermore, this phenomenological approach does not provide a rigorous scheme to consistently derive two- and many-body forces and compatible electroweak currents. In addition, there is no clear way to properly assess the theoretical uncertainty associated with the nuclear potentials and currents.

These shortcomings were addressed when a new phase in the evolution of microscopic models began in the early 1990's with the emergence of $\chi$EFT~\cite{Weinberg:1990rz,Weinberg:1991um,Weinberg:1992yk}. 
$\chi$EFT is a low-energy effective theory of QCD and provides the most general scheme accommodating all possible interactions among nucleons and pions ($\Delta$-less $\chi$EFT) compatible
with the relevant symmetries and symmetry breakings -- in particular chiral symmetry -- of low-energy QCD. In some modern approaches, the choice of degrees of freedom also includes the $\Delta$ isobar ($\Delta$-full $\chi$EFT), because the $\Delta$-nucleon mass splitting is only $300\,\rm MeV\sim 2 m_{\pi}$.

By its own nature, the $\chi$EFT formulation has an expansion in powers of pion momenta as its organizing principle. Most chiral interactions employed in recent nuclear structure and reaction calculations are based on Weinberg power counting. Within Weinberg power counting, the interactions are expanded in powers of the typical momentum $p$ over the breakdown scale $\Lambda_b\sim\rm GeV$, $Q=p/\Lambda_b$, where the breakdown scale denotes momenta at which the short distance structure becomes important and cannot be neglected and absorbed into contact interactions anymore (see references~\cite{Epelbaum:2008ga,Machleidt:2011zz,Machleidt:2016rvv,Machleidt:2017vls} for recent review articles). 
It is important mentioning that alternative power-counting schemes have been also suggested~\cite{Kaplan:1998tg,Kaplan:1998we,Nogga:2005hy,PavonValderrama:2005wv,Long:2011xw,vanKolck:1994yi} but not fully explored.

This expansion introduces an order by order scheme, defined by the power $\nu$ of the expansion scale $Q$ associated with each interaction terms: leading order (LO) for $\nu=0$, next-to-leading order (NLO) for $\nu=2$, next-to-next-to-leading order (N$^2$LO) for $\nu=3$ and so on. Similarly as for nuclear interactions, such a scheme can also be developed for electroweak currents. Therefore, $\chi$EFT provides a rigorous scheme to systematically construct many-body forces and consistent electroweak currents, and tools to estimate their uncertainties~\cite{Furnstahl:2014xsa,Epelbaum:2014efa,Furnstahl:2015rha,Wesolowski:2015fqa,Melendez:2017phj,Wesolowski:2018lzj}. 

Chiral nuclear forces are comprised of both pion-exchange contributions 
and contact terms. Pion-exchange contributions represent the long-range part of nuclear interactions and some of the pion-nucleon ($\pi N$) couplings entering sub-leading diagrams can consistently be obtained from low-energy $\pi N$ scattering data~\cite{Krebs:2007rh,Ditsche:2012fv,Hoferichter:2015dsa,Hoferichter:2015hva,Hoferichter:2015tha,Siemens:2016hdi,Siemens:2016jwj,Yao:2016vbz}. On the other end, contact terms encode the short-range physics, and their strength is specified by unknown low-energy constants (LECs), obtained by fitting experimental data. Similarly to the phenomenological interactions, the LECs entering the $N\!N$ component are obtained by fitting $N\!N$ scattering data up to $300\,\rm MeV$ lab energies, while the LECs involved in the $3N$ terms are fixed by reproducing properties of light-nuclei. This optimization procedure involves separate fit of the $N\!N$ and $3N$ terms. Recently, a different strategy has been introduced by Ekstr\"om \textit{et al.}~\cite{Ekstrom:2015rta}.
This new approach is based on a simultaneous fit of the $N\!N$ and $3N$ forces to low-energy $N\!N$ data, deuteron binding energy, and binding energies and charge radii of hydrogen, helium, carbon, and oxygen isotopes.

Within $\chi$EFT, many studies have been carried out dealing with the construction and optimization of $N\!N$ and $3N$ interactions~\cite{Ordonez:1995rz,Kaiser:1997mw,Kaiser:1998wa,Epelbaum:1998ka,Epelbaum:1999dj,Kaiser:1999ff,Kaiser:1999jg,Kaiser:2001dm,Kaiser:2001pc,Kaiser:2001at,Entem:2003ft,Machleidt:2011zz,Krebs:2007rh,vanKolck:1994yi,Epelbaum:2002vt,Navratil:2007zn,Bernard:2011zr,Girlanda:2011fh,Krebs:2012yv,Epelbaum:2014efa,Epelbaum:2015,Entem:2014msa,Entem:2015xwa,Ekstrom:2013kea,Ekstrom:2015rta,Ekstrom:2017koy,Reinert:2017usi,Binder:2018pgl,Krebs:2018jkc} and accompanying isospin-symmetry-breaking corrections~\cite{Friar:1999zr,Friar:2004ca,Friar:2004rg}.
These interactions are typically formulated in momentum space, and include cutoff functions to regularize their behavior at large momenta. This makes them strongly non-local when Fourier-transformed in configuration space, and therefore unsuitable for use with QMC methods.
In this context, an interaction is local if it depends solely on the momentum transfer $\textbf{q}={\textbf p} -{\textbf p}^\prime$ (${\textbf p}$ and ${\textbf p}^\prime$ are the initial and final relative momenta of the two nucleons), which, upon Fourier transform, leads to dependencies solely on $\textbf{r}$.
However, interactions in momentum-space can also depend on the momentum scale ${\textbf k}=({\textbf p}^\prime +{\textbf p})/2$, whose Fourier transform introduces derivatives in coordinate space. 
These ${\textbf k}$ dependencies, and thus non-localities, come about because of (i) the specific functional choice made to regularize the momentum space potentials in terms of the two momentum scales $\textbf{p}$ and $\textbf{p}'$, and (ii) contact interactions that explicitly depend on $\textbf{k}$.

In recent years, local configuration-space chiral $N\!N$ interactions have been derived by two groups. On the one side, the authors of references~\cite{Gezerlis:2013ipa,Gezerlis:2014zia} constructed $N\!N$ local chiral potentials within $\Delta$-less $\chi$EFT by including one- and two-pion exchange contributions and contact terms up to N$^2$LO in the chiral expansion. The contact terms are regularized in coordinate space by a cutoff function depending only on the relative distance between the two nucleons, and use Fierz identities~\cite{Fierz1937} to remove completely the dependence on the relative momentum of the two nucleons, by selecting appropriate combinations of contact operators. Their strength is characterized by 11 LECs, fixed by performing order by order $\chi^2$ fit to $N\!N$ phase shifts from the Nijmegen PWA up to $150\,\rm MeV$ lab energy. The fitting procedure is carried out for different values of the cutoff $R_0$ in the range of $R_0 = 1.0-1.2\,\rm fm$.
The motivations why the authors of references~\cite{Gezerlis:2013ipa,Gezerlis:2014zia} truncated the chiral expansion of these local potentials at N$^2$LO is because at this order it is i) possible to have a fully local representation of the $N\!N$ chiral interactions and ii) the inclusion of consistent $3N$ force is straightforward. In their models, the unknown $3N$ LECs are obtained by reproducing the binding energy of \isotope[4]{He} as well as the $P$-wave $n-\alpha$ elastic scattering phase shifts. In addition, they explore different parametrization for the $3N$ force, accordingly to Fierz identities~\cite{Lynn:2015jua,Lynn:2017fxg,Lonardoni:2018prc}.
In the present work, we are referring to a set of these local chiral interactions, specifically the $(D2,E\tau)$ model with $R_0=1.0\,\rm fm$ of reference~\cite{Lonardoni:2018prc}, as GT+E$\tau$-1.0.

On the other side, the authors of references~\cite{Piarulli:2014bda,Piarulli:2016vel} developed a different set of $N\!N$ local chiral interactions by i) including diagrams with the virtual excitation of $\Delta$-isobars in the TPE contributions up to N$^2$LO ($\Delta$-full $\chi$EFT), ii) retaining contact terms up to N$^3$LO. The LECs entering the $N\!N$ contact interactions in this model are constrained to reproduce $N\!N$ scattering data
from the most recent and up-to-date database collected by the Granada 
group~\cite{Perez:2013jpa,Perez:2013oba,Perez:2014yla}. The contact terms are implemented via a Gaussian representation
of the three-dimensional delta function with $R_S$ as the Gaussian 
parameter~\cite{Piarulli:2016vel,Piarulli:2014bda,Baroni:2018fdn}. 
The pion-range operators are regularized at
high-value of momentum transfer via a special radial function characterized
by the cutoff $R_L$~\cite{Piarulli:2016vel,Piarulli:2014bda,Baroni:2018fdn}.
There are two classes of these potentials. Class I (II) are fit
to data up to $125\,\rm MeV$ ($200\,\rm MeV$). For each class, two combinations
of short- and long-range regulators have been used, namely
$(R_S, R_L)=(0.8, 1.2)\,\rm fm$ (models NV2-Ia and NV2-IIa) and 
$(R_S, R_L)=(0.7, 1.0)\,\rm fm$ (models NV2-Ib and NV2-IIb).
Class I (II) fits about 2700 (3700)
data points with a $\chi^2/{\rm datum}\lesssim 1.1$ 
($\lesssim 1.4$)~\cite{Piarulli:2016vel,Piarulli:2014bda}. 
In conjunction with these models, two distinct sets of $\Delta$-full $3N$ interactions have
also been constructed up to N$^2$LO. In the first, the $3N$ unknown LECs were determined by simultaneously reproducing the experimental trinucleon ground-state energies and neutron-deuteron ($nd$) doublet scattering length for each of the $NN$ models considered, namely NV2-Ia/b and NV2-IIa/b~\cite{Piarulli:2017dwd,Baroni:2016xll}. In the second set, these LECs were
constrained by fitting, in addition to the trinucleon energies, the empirical value of the Gamow-Teller matrix element
in tritium $\beta$-decay~\cite{Baroni:2018fdn}. The resulting Hamiltonians were labelled as NV2+3-Ia/b and NV2+3-IIa/b (or
Ia/b and IIa/b for short) in the first case, and as NV2+3-Ia$^*$/b$^*$ and NV2+3-IIa$^*$/b$^*$ (or Ia$^*$/b$^*$ and IIa$^*$/b$^*$) in the second.

The interactions between external electroweak probes – electrons and neutrinos – and interacting nuclear systems is described by a set of effective nuclear currents and charge operators. Analogously to the nuclear interactions, electroweak currents can also be expressed as
an expansion in many-body operators that act on nucleonic degrees of freedom. Electroweak currents have been developed in both meson-exchange and $\chi$EFT approaches. We refrain to discuss them in this work, redirecting the interested reader to dedicated reviews~\cite{Carlson:2015,Bacca:2014tla,Marcucci:2015rca,Lynn:2019rdt} and references therein.

\section{Quantum Monte Carlo methods} 
The $\chi$EFT Hamiltonians and the consistent electroweak currents discussed in the previous Section are the main input of sophisticated many-body methods aimed at solving with controlled approximations the nuclear many-body Schr\"odinger equation 
\begin{align}
H|\Psi_n\rangle = E_n |\Psi_n\rangle\, .
\end{align}
This is a highly non-trivial problem, mainly because of the non-perturbative nature and the strong spin-isospin dependence of realistic nuclear forces. In this work, we will focus on QMC techniques, namely the variational Monte Carlo (VMC), the Green's function Monte Carlo (GFMC), and the auxiliary-field diffusion Monte Carlo (AFDMC) methods.

\subsection{Variational Monte Carlo}
The variational Monte Carlo method is routinely used to obtain approximate solutions to the many-body Schr\"odinger equation for a wide range of strongly interacting nuclear systems, including few-body nuclei, light closed-shell nuclei, and nuclear and neutron matter~\cite{Carlson:2015}. The VMC algorithm relies on the Rayleigh-Ritz variational principle
\begin{align}
\frac{\langle \Psi_T | H | \Psi_T \rangle}{\langle \Psi_T | \Psi_T \rangle} = E_T \geq E_0 
\label{eq:H_exp}
\end{align}
to find the optimal set of variational parameters defining the trial wave function $\Psi_T$. 
As far as the nuclear many-body problem is concerned, it is customary to assume that the trial state factorizes into long- and short-range components
\begin{align}
|\Psi_T \rangle = \Big(1-\sum_{i<j<k} F_{ijk}\Big) \Big(\mathcal{S} \prod_{i<j} F_{ij} \Big) | \Phi_J\rangle\, ,
\label{eq:psi_T}
\end{align}
where $F_{ij}$ and $F_{ijk}$ are two- and three-body correlations, respectively.
The symbol $\mathcal{S}$ indicates a symmetrized product over nucleon pairs since, in general, the $F_{ij}$ do not commute. VMC calculations explicitly account for the underlying strong alpha-cluster structure of light nuclei. For instance, the totally antisymmetric Jastrow wave function of $p$-shell nuclei is constructed from a sum over independent-particle terms, $\Phi_A$, each having four nucleons in an $\alpha$-like core and the remaining $(A-4)$ nucleons in $p$-shell orbitals~\cite{Pieper:2002ne}:
\begin{align}
| \Phi_J\rangle=&\mathcal{A}\left[\prod_{i<j<k} f^{c}_{ijk} \prod_{i<j\leq 4} f_{ss}(r_{ij}) \prod_{k\leq 4 < l \leq A} f_{sp}(r_{kl})\right. \nonumber\\
&\left.\times \sum_{LS[n]} \Big(\beta_{LS[n]} \prod_{4<l<m\leq A} f_{pp}^{[n]}(r_{lm})\, | \Phi_A(LS[n]J J_z T_z)_{1234:5\dots A}\rangle\Big)\right]\,.
\label{def:phiJ_gfmc}
\end{align}
The operator $\mathcal A$ stands for an antisymmetric sum over all possible $\binom{A}{4}$ partitions of the $A$ particles into four $s$-shell and $(A-4)$ $p$-shell states. As suggested by standard shell-model studies, the independent-particle wave function $| \Phi_A(LS[n]J J_z T_z)_{1234:5\dots A}\rangle$ with the desired $J M$ value of a given nuclear state is obtained using $LS$ coupling, which is most efficient for nuclei with up to $A=12$. The symbol $[n]$ is the Young pattern that indicates the spatial symmetry of the angular momentum coupling of the $p$-shell nucleons~\cite{Pudliner:1997ck}. Note that $|\Phi_A(LS[n]J J_z T_z)_{1234:5\dots A}\rangle$ is chosen to be independent of the center of mass as it is expressed in terms of the intrinsic coordinates
\begin{align}
    \mathbf{r}_i \to \mathbf{r}_i - \mathbf{R}_{\rm CM}\,, \qquad \, \mathbf{R}_{\rm CM}=\frac{1}{A} \sum_{i=1}^A \mathbf{r}_i\,. 
    \label{eq:r_intr}
\end{align}
The pair correlation for particles within the $s$-shell, $f_{ss}$, arises from the structure of the $\alpha$ particle. The $f_{sp}$ is similar to the $f_{ss}$ at short range, but with a long-range tail that goes to unity at large distances, allowing the wave function to develop a cluster structure. Finally, $f_{pp}$ is set to give the appropriate cluster structure outside the $\alpha$ core. The three-body central correlations, induced by the two-body potential has the following operator independent form
\begin{align}
f^c_{ijk}=1-q_1^c(\mathbf{r}_{ij}\cdot \mathbf{r}_{ik})(\mathbf{r}_{ij}\cdot \mathbf{r}_{jk})(\mathbf{r}_{ik}\cdot \mathbf{r}_{jk})e^{-q_2^c(r_{ij}+r_{ik}+r_{jk})}\,,
\label{eq:fc_ijk}
\end{align}
where $q_1^c$ and $q_2^c$ are variational parameters. In addition the the scalar correlations of Equation~(\ref{def:phiJ_gfmc}), VMC trial wave functions include spin-dependent nuclear correlations, whose operator structure reflects the one of the $N\!N$ potential of Equation~(\ref{eq:NN_pot})
\begin{align}
F_{ij}=\left(1+U_{ij}\right)=\Big(1+\sum_{p=2}^6 u^p(r_{ij})\mathcal O^p_{ij}\Big)\, .
\label{eq:Fij}
\end{align}
More sophisticated trial wave functions can be constructed by explicitly accounting for spin-orbit correlations, as, for instance, in the cluster variational Monte Carlo calculations of reference~\cite{Lonardoni:2017}. However, the computational cost of these additional terms is significant, while the gain in the variational energy is relatively small~\cite{Wiringa:2000gb}.
The radial functions $u^p(r_{ij})$ are generated by minimizing the two-body cluster energy of the interaction $\bar{v}-\lambda$, with
\begin{align}
\bar{v}-\lambda = \sum_{p=1}^{18} \left(\alpha_p v^p(r_{ij}) \mathcal O^p_{ij} - \lambda_p(r_{ij})\right)\, .
\end{align}
The variational parameters $\alpha_p$ simulate the quenching of spin-isospin interactions between particles $i$ and $j$ due to interactions of these particles with others in the system. The Lagrange multipliers $\lambda_p(r_{ij})$ account for short-range screening effects, and are fixed at large distances by the asymptotic behavior of the correlation functions, which is encoded by an additional set of variational parameters. 
The quality of the trial wave function is improved by reducing the strength of the spin- and isospin-dependent correlation functions $u^p(r_{ij})$ when a particle $k$ comes close to the pair $ij$~\cite{Lomnitz-Adler:1981dmh}
\begin{align}
u^p(r_{ij})\to \left[\prod_{k\neq i\neq j} f^p_{ijk}(\mathbf{r}_{ij},\mathbf{r}_{ik})\right] u^p(r_{ij})\, ,
\end{align}
where the three-body operator-dependent correlation induced by the $N\!N$ interaction is usually expressed as
\begin{align}
f^p_{ijk}(\mathbf{r}_{ij},\mathbf{r}_{ik})=1-q_1^p(1-\hat{\mathbf{r}}_{ik}\cdot \hat{\mathbf{r}}_{jk})e^{-q_2^p(r_{ij}+r_{ik}+r_{jk})}\, ,
\end{align}
with $q_1^p$ and $q_2^p$ being variational parameters~\cite{Pudliner:1997ck}. 
The three-body correlation operator $F_{ijk}$ turns out to be particularly relevant for when $3N$ interactions are present in the nuclear Hamiltonian. In this case, its form is suggested by perturbation theory
\begin{align}
F_{ijk} = \sum_q \epsilon_q V_{ijk}^q (y_q r_{ij}, y_q r_{ik},y_q r_{jk})\, ,
\label{eq:Fijk}
\end{align}
where $y_q$ is a scaling parameter, and $\epsilon_q$ a small constant. The superscript $q$ indicates the various terms of the $3N$ force. It has been shown that the vast majority of the $3N$ correlations can be recovered by omitting the commutator term $\epsilon_C V_{ijk}^C$, provided that the strength of the anticommutator term $\epsilon_A$ is opportunely adjusted. This allows to save a significant amount of computing time, since anticommutators involving pairs $ij$ and $jk$ can be expressed as generalized tensor operators involving the spins of nucleons $i$ and $k$ only. Hence, the computing time scales as the number of pairs rather than the number of triplets~\cite{Pudliner:1997ck}.

The expectation values of the form of Equation~(\ref{eq:H_exp}) contain multi-dimensional integrals over all particle positions
\begin{align}
\langle \mathcal O \rangle = \frac{\int d\mathbf{R} \Psi^\dagger_T(\mathbf{R}) \mathcal O \Psi_T(\mathbf{R}) }{\int d\mathbf{R}\Psi^\dagger_T(\mathbf{R}) \Psi_T(\mathbf{R})}\,.
\label{eq:expectation}
\end{align}
A deterministic integration of the above integral is computationally prohibitive, therefore Metropolis Monte Carlo techniques are employed to stochastically evaluate it. The order of operators in the symmetrized product of Equation~(\ref{eq:psi_T}), denoted by $p$ and $q$ for the left and right hand side wave functions, respectively, is also sampled. The $3A$-dimensional integration is facilitated by introducing a probability distribution, $W_{pq}(\mathbf{R})$, such that
\begin{align}
\langle \mathcal O \rangle =  \frac{\sum_{p,q}\int d\mathbf{R} \frac{\Psi^\dagger_{T,p}(\mathbf{R}) \mathcal O  \Psi_{T,q}(\mathbf{R})}{W_{pq}(\mathbf{R})} W_{pq}(\mathbf{R})}{\sum_{p,q}\int d\mathbf{R} \frac{\Psi^\dagger_{T,p}(\mathbf{R}) \Psi_{T,q}(\mathbf{R})}{W_{pq}(\mathbf{R})} W_{pq}(\mathbf{R})}\,.
\end{align}
In standard VMC calculations, one usually takes $W_{pq}(\mathbf{R})=|\Re(\Psi^\dagger_{T,p}(\mathbf{R}) \Psi_{T,q}(\mathbf{R}))|$, even though simpler choices might be used to reduce the computational cost. The Metropolis algorithm is used to stochastically sample the probability distribution $W_{pq}(\mathbf{R})$ and obtain a collection of uncorrelated or independent configurations. 

Since the nuclear interaction is spin-isospin dependent, the trial state is a sum of complex amplitudes for each spin-isospin state of the system
\begin{align}
|\Psi_T \rangle = \sum_{i_s\leq n_s,i_t\leq n_t} a(i_s,i_t;\mathbf{R}) | \chi_{i_s}\, \chi_{i_t} \rangle \, .
\label{eq:spin_isospin}
\end{align}
The $n_s=2^A$ many-body spin states can be written as
\begin{align}
|\chi_1\rangle &= |\downarrow_1,\downarrow_2,\dots,\downarrow_A\rangle \nonumber\\
|\chi_2\rangle &= |\uparrow_1,\downarrow_2,\dots,\downarrow_A\rangle \nonumber\\
|\chi_3\rangle &= |\downarrow_1,\uparrow_2,\dots,\downarrow_A\rangle \nonumber\\
\dots \nonumber\\
|\chi_{n_s}\rangle &= |\uparrow_1,\uparrow_2,\dots,\uparrow_A\rangle \, 
\end{align}
and the isospin ones can be recovered by replacing $\downarrow$ with $n$ and $\uparrow$ with $p$. Note that, because of charge conservation, the number of isospin states reduces to $n_t=\binom{A}{Z}$. To construct the trial state, one starts from the mean-field component $|\Phi_A(LS[n]J J_z T_z)_{1234:5\dots A}\rangle$. For fixed spatial coordinates $\mathbf{R}$, the spin-isospin independent correlations needed to retrieve $| \Phi_J\rangle$ are simple multiplicative factors, common to all spin amplitudes. The symmetrized product of pair correlation operators is evaluated by successive operations for each pair, sampling their ordering as alluded to earlier. 
As an example, consider the application of the operator ${\bm \sigma}_1 \cdot {\bm \sigma}_2$ on a three-body spin state (for simplicity we neglect the isospin components). Noting that ${\bm \sigma}_i \cdot {\bm \sigma}_j=2\,\mathcal P_{ij}^\sigma -1$, where $2\,\mathcal P_{ij}^\sigma$ exchanges the spin of particles $i$ and $j$, we obtain:
\begin{align}
{\bm \sigma}_1 \cdot {\bm \sigma}_2
\left(
\begin{array}{c}
a_{\uparrow\uparrow\uparrow}\\
a_{\uparrow\uparrow\downarrow}\\
a_{\uparrow\downarrow\uparrow}\\
a_{\uparrow\downarrow\downarrow}\\
a_{\downarrow\uparrow\uparrow}\\
a_{\downarrow\uparrow\downarrow}\\
a_{\downarrow\downarrow\uparrow}\\
a_{\downarrow\downarrow\downarrow}
\end{array} \right)
 =  \left(
\begin{array}{c}
 a_{\uparrow\uparrow\uparrow}\\
 a_{\uparrow\uparrow\downarrow}\\
 2a_{\downarrow\uparrow\uparrow}-a_{\uparrow\downarrow\uparrow}\\
 2a_{\downarrow\uparrow\downarrow}-a_{\uparrow\downarrow\downarrow}\\
 2a_{\uparrow\downarrow\uparrow}-a_{\downarrow\uparrow\uparrow}\\
 2a_{\uparrow\downarrow\downarrow}-a_{\downarrow\uparrow\downarrow}\\
 a_{\downarrow\downarrow\uparrow}\\
 a_{\downarrow\downarrow\downarrow}
\end{array} \right)\,.
\end{align}
Hence, the many-body spin-isospin basis is closed under the action of the operators contained in the nuclear Hamiltonian. 

Most of the computing time is spent on spin-isospin operations like the one just described. They amount to an iterative sequence of large sparse complex matrix multiplications that are performed on-the-fly using explicitly coded subroutines, which mainly rely on three useful matrices. The first matrix $m(i,i_s)$ gives the $z$-component of the spin of particle $i$ associated to the many-body spin-state $i_s$. A second useful matrix is $n_{\rm exch} (k_{ij}, i_s)$, that provides the number of the many-body spin state obtained by exchanging the spins of particles $i$ and $j$, belonging to the pair labeled $k_{ij}$ in the state $i_s$. The matrix $n_{\rm flip}(i,i_s)$ yields the number of the spin state obtained by flipping the spin of particle $i$ in the spin state. The action of the operator ${\bm \sigma}_1 \cdot {\bm \sigma}_2$ can then be expressed as
\begin{align}
{\bm \sigma}_1 \cdot {\bm \sigma}_2 \, \sum_{i_s,i_t} a(i_s,i_t;\mathbf{R}) | \chi_{i_s}\, \chi_{i_t} \rangle =
\sum_{i_s,i_t} \big[2 a(i_s,i_t;\mathbf{R}) - a(n_{\rm exch} (k_{ij}, i_s),i_t;\mathbf{R})\big]| \chi_{i_s}\, \chi_{i_t} \rangle \,.
\end{align}
By utilizing this representation, we only need to evaluate $2^A$ operations for each pair, instead of the $2^A \times 2^A$ operations that are required using a simple matrix representation in spin space. The tensor operator is slightly more complicated to evaluate and requires both matrices $m(i,i_s)$ and $n_{\rm flip}(i,i_s)$~\cite{Pudliner:1996}. Analogous matrices are employed to perform operations on the isospin space, as the two representations are practically identical.

The expectation values of Equation~(\ref{eq:expectation}) are evaluated by having the operators act entirely on the right hand side of the trial wave function. The matrix machinery used to apply the spin-dependent correlation operators is also used to evaluate $\mathcal O |\Psi_{T,p}\rangle$. A simple scalar product of this quantity with $\langle \Psi_{T,q}|$, provides the numerator of the local estimate $\Psi_{T,q}^\dagger (\mathbf{R}) \mathcal O  \Psi_{T,p}(\mathbf{R}) / W_{pq}(\mathbf{R})$ and $W_{pq}(\mathbf{R})$ is computed in a similar fashion. The first and second derivatives of the wave function are numerically computed by means of the two- and three-point stencil, respectively. Hence, to determine the kinetic energy, $6A+1$ evaluations of $\Psi_{T}(\mathbf{R})$ are needed. Finally, using the trick described in reference~\cite{Schiavilla:1985gb}, we can evaluate the action of the angular momentum dependent terms in the potential evaluating $\Psi_{T}(\mathbf{R})$ an additional $3A(A-1)/2$ times. 

Not only does the size of the wave vector grows exponentially with the number of nucleons, but so does the number of evaluations necessary to calculate the energy, limiting the applicability of the VMC method to $A\leq 12$ nuclei. Sampling the spin-isospin state and evaluating the trial wave function's amplitude for that sampled state still requires a number of operations exponential in the particle number, bringing little savings in terms of computing time. Extending VMC calculations to larger nuclear systems requires devising trial wave functions that can capture most of physics of the system while requiring computational time that scales polynomially with $A$.

\subsection{Green's function Monte Carlo}
Green's function Monte Carlo overcomes the limitations intrinsic to the variational ansatz by using an imaginary-time projection technique to enhance the true ground-state component of a starting trial wave function
\begin{align}
\ket{\Psi_0}\propto\lim_{\tau\to\infty}e^{-(H-E_T)\tau}\ket{\Psi_T}\,.
\end{align}
In the above equation, $\tau$ is the imaginary time, and $E_T$ is a parameter used to control the normalization. In addition to ground-states properties, excited states can be computed within GFMC. The imaginary-time diffusion yields the lowest-energy eigenstate with the same quantum numbers as $|\Psi_T\rangle$. Thus, to obtain an excited state with distinct quantum numbers from the ground state, one only needs to construct a trial wave function with the appropriate quantum numbers. If the excited-state quantum numbers coincide with those of the ground state, more care is needed, but precise results for such states can still be obtained~\cite{Carlsson:2015vda}.

Except for some specific cases, the direct computation of the propagator $e^{-H\tau}$ for arbitrary values of $\tau$ is typically not possible. For small imaginary times $\delta\tau=\tau/N$ with $N$ large, the calculation is tractable, and the full propagation to large imaginary times $\tau$ can be recovered through the following path integral
\begin{align}
\Psi(\tau, \mathbf{R}_N ) =\int\prod_{i=0}^{N-1}\dd\vb{R}_i\matrixel{\vb{R}_N}{e^{-(H-E_T)\delta\tau}}{\vb{R}_{N-1}}\cdots\matrixel{\vb{R}_1}{e^{-(H-E_T)\delta\tau}}{\vb{R}_0}\!\braket{\vb{R}_0}{\Psi_T}\,.
\label{eq:path}
\end{align}
The GFMC wave function at imaginary time $\tau+\delta\tau$ can be written in an integral form  
\begin{align}
\Psi(\tau + \delta\tau, \mathbf{R}_{i+1} ) = \int d \mathbf{R}_iG_{\delta\tau}(\vb{R}_{i+1},\vb{R}_{i})\Psi(\tau, \mathbf{R}_i )\, ,
\label{eq:prop}
\end{align}
where we defined the short-time propagator, or Green's function,
\begin{align}
G_{\delta\tau}(\vb{R}_{i+1},\vb{R}_{i})=\matrixel{\vb{R}_{i+1}}{e^{-H\delta\tau}}{\vb{R}_{i}}\, .
\end{align}
Monte Carlo techniques are used to sample the paths by simultaneously evolving a set of configurations -- dubbed {\it walkers} -- in imaginary time, until the distribution converges to the ground-state wave function~\cite{Foulkes:2001zz}. To avoid the large statistical errors arising from configurations that diffuse into regions where they make very little contribution to the ground-state wave function, the diffusion process is guided by introducing an importance-sampling function $\Psi_I(\mathbf{R})$, which has the same quantum numbers as the ground-state. The importance function is typically taken to coincide with the variational wave function, but different choices are possible. Multiplying \cref{eq:prop} on the left by $\Psi_I^\dagger(\mathbf{R}_{i+1})$ yields
\begin{align}
\Psi_I^\dagger(\mathbf{R}_{i+1}) \Psi(\tau + \delta\tau, \mathbf{R}_{i+1} ) = \int d \mathbf{R}_i \left[ \Psi_I^\dagger(\mathbf{R}_{i+1}) G_{\delta\tau}(\vb{R}_{i+1},\vb{R}_{i})  \frac{1}{\Psi_I^\dagger(\mathbf{R}_{i})} \right]\Psi_I^\dagger(\mathbf{R}_{i}) \Psi(\tau, \mathbf{R}_i )\, .
\label{eq:prop2}
\end{align}
The quantity within squared brackets is the importance-sampled propagator $G_{\delta\tau}^I(\vb{R}_{i+1},\vb{R}_{i})$. Note that a set of walkers can be sampled from $\Psi_I^\dagger(\mathbf{R}_{i}) \Psi(\tau + \delta\tau, \mathbf{R}_{i} )$ only if this density is positive definite. In this case, the latter can be interpreted as a probability density distribution and its integral determines the size of the population, i.e., the number of walkers. In Fermion systems, however, the positiveness of $\Psi_I^\dagger(\mathbf{R}_{i}) \Psi(\tau + \delta\tau, \mathbf{R}_{i} )$ is only granted for exact importance-sampling functions. In general, the nodal surface of the ground state can be different from that of $\Psi_I$. We will return to this point later on. The importance function can be expanded in terms of eigenstates of the Hamiltonian as
\begin{align}
\Psi_I(\vb{R}_{i}) = \sum_n c_n \Psi_n(\vb{R})\,.
\end{align}
The Green's function can also be expressed in terms of the same eigenstates:
\begin{align}
G_{\delta\tau}(\vb{R}_{i+1},\vb{R}_{i})= \sum_n \Psi_n(\mathbf{R}_{i+1}) e^{-(E_n-E_T)\delta\tau} \Psi_n^\dagger(\mathbf{R})\,.
\end{align}
Inserting the last two relations into Equation~(\ref{eq:prop}) and integrating over $\mathbf{R}_{i+1}$, we get
\begin{align}
\sum_n c_n^* \int d\vb{R}_{i+1} \Psi_n^\dagger(\vb{R}_{i+1}) \Psi(\tau + \delta\tau, \mathbf{R}_{i+1} )= \sum_n c_n^* \int d\vb{R}_i \Psi_n^\dagger(\vb{R}_i)\, e^{-(E_n-E_T)\delta\tau}\Psi(\tau, \mathbf{R}_i )\,.
\end{align}
If the importance-sampling function closely resembles the ground-state wave function, then $c_n^*\simeq \delta_{n0}$ and $E_T\simeq E_0$, implying
\begin{align}
\int d\vb{R}_{i+1} \Psi_0^\dagger(\vb{R}_{i+1}) \Psi(\tau + \delta\tau, \mathbf{R}_{i+1} )\simeq \int d\vb{R}_i \Psi_0^\dagger(\vb{R}_i)\Psi(\tau, \mathbf{R}_i )\,.
\end{align}
Therefore, having accurate importance function reduces the fluctuations in the population size from one time step to the next, thereby reducing the statistical errors in the calculation. 

A common approximation for the short-time propagator is based upon the Trotter-Suzuki expansion
\begin{align}
\label{eq:trotter}
\begin{split}
G_{\delta\tau}(\vb{R}_{i+1},\vb{R}_i)
&=e^{-V(\vb{R}_{i+1})\delta\tau/2} \langle \vb{R}_{i+1}| e^{-T\delta\tau} |\vb{R}_i\rangle e^{-V(\vb{R}_i)\delta\tau/2}+o(\delta\tau^3)\,.
\end{split}
\end{align}
Here, $T$ is the kinetic energy giving rise to the free-particle propagator that, for non-relativistic systems, can be expressed as a simple Gaussian in configuration space
\begin{align}
 \langle \vb{R}_{i+1}|e^{-T\delta\tau}|\vb{R}_i\rangle=G_{\delta\tau}^0(\vb{R}_{i+1},\vb{R}_{i})=\left[\frac{1}{\lambda^3 \pi^{3/2}}\right]^A e^{-(\vb{R}_{i+1}-\vb{R}_i)^2/\lambda^2},
 \label{eq:prop_free}
\end{align}
with $\lambda^2=4\frac{\hbar^2}{2m}\delta\tau$. The exponentials of the two-body potentials can be approximated to first order by turning the sums over pairs in the exponent into a symmetrized product of exponentials of the individual pair potentials. The first six terms of the potential can be easily exponentiated, while momentum dependent terms cannot be treated this way. A simple way to include them consists in expanding the exponential of the momentum dependent terms to first order in $\delta\tau$ and use integration by parts to let the derivatives act on the free-particle Green's function. This approach can only be successfully applied to the terms in the potential that are linear in momentum, such as $\mathbf{L}\cdot\mathbf{S}$ and $(\mathbf{L}\cdot\mathbf{S})\,\bm\tau_i\cdot\bm\tau_j$~\cite{Carlson:1988zz}. On the other hand, contributions to the potential that are quadratic in the momentum cannot be evaluated to first order in this manner. For this reason we use approximations to the full $N\!N$ potentials, such as the AV8$^\prime$ interaction, that only contain the first eight operators. The difference between AV18 and AV8$^\prime$ is treated in perturbation theory.

More sophisticated alternatives of reducing the time-step error exist and are routinely used in GFMC calculations. The most common one consists in building the Green's function operator as a product of exact two-body propagators
\begin{align}
G_{\delta\tau}(\vb{R}_{i+1},\vb{R}_i)=\left(\mathcal{S}\prod_{j<k} \frac{g_{jk}(\mathbf{r}_{jk,\,i},\mathbf{r}_{jk,\,i+1})}{g^0_{jk}(\mathbf{r}_{jk,\,i},\mathbf{r}_{jk,\,i+1})}\right) G_{\delta\tau}^0(\vb{R}_{i+1},\vb{R}_{i})\,,
\end{align}
where $g_{jk}(\mathbf{r}_{jk,\,i},\mathbf{r}_{jk,\,i+1})$ is the exact two-body propagator and $g^0_{jk}(\mathbf{r}_{jk,\,i},\mathbf{r}_{jk,\,i+1})$ is the two-body free- particle propagator~\cite{Schmidt:1995}. At variance with the propagator of \cref{eq:trotter}, terms quadratic in the angular momentum can in principle be accounted for into the exact pair propagator. However, the inclusion of these terms requires the sampled distribution to have the same locality structure to keep statistical errors under control. Thus, simplified AV8$^\prime$ potentials are also used in the pair propagator, even though in this case no approximations in treating $\mathbf{L}\cdot\mathbf{S}$ and $(\mathbf{L}\cdot\mathbf{S})\,\bm\tau_i\cdot\bm\tau_j$ terms are necessary.

Since the matrix $V$ is the spin/isospin-dependent interaction, the propagator is in turn a matrix in spin-isospin space. To deal with it, first a scalar approximation to the importance sampled Green's function, denoted as $\tilde{G}_{\delta\tau}^I(\vb{R}_{i+1},\vb{R}_{i})$, is introduced. Recalling the form of the importance sampled Green's function 
\begin{align}
G_{\delta\tau}^I(\vb{R}_{i+1},\vb{R}_{i})= \frac{\Psi_I^\dagger(\mathbf{R}_{i+1})}{\Psi_I^\dagger(\mathbf{R}_{i})} G_{\delta\tau}(\vb{R}_{i+1},\vb{R}_{i})\,,
\label{eq:prop_I}
\end{align}
constructing its scalar counterpart requires defining a scalar approximation for the importance-sampling function, which can be taken to be $\tilde{\Psi}_I(\vb{R})=\sqrt{\Psi_J^\dagger(\vb{R}) \Psi_J(\vb{R})}$. As for the potential, we can use the average of the central parts in the $^1S_0$ and $^3S_1$ channels, thus
\begin{align}
\tilde{G}_{\delta\tau}^I(\vb{R}_{i+1},\vb{R}_{i})=\frac{\tilde{\Psi}_I(\vb{R}_{i+1})}{\tilde{\Psi}_I(\vb{R}_i)} e^{-[V_{10}(\mathbf{R}_{i+1}) + V_{01}(\mathbf{R}_{i+1})]\delta\tau/4}G_{\delta\tau}^0(\vb{R}_{i+1},\vb{R}_{i})
e^{-[V_{10}(\mathbf{R}_i) + V_{01}(\mathbf{R}_i]\delta\tau/4}\,.
\end{align}
At each time-step, the walkers are propagated with $G_{\delta\tau}^0(\vb{R}_{i+1},\vb{R}_{i})$ by sampling a $3A$-dimensional vector from a gaussian distribution to shift the spatial coordinates. To remove the linear terms coming from the exponential of \cref{eq:prop_free}, we use two mirror points $\mathbf{R}_{i+1}=\mathbf{R}_i \pm \delta \mathbf{R}$ and we consider the corresponding two weights
\begin{align}
w_\pm = \frac{\tilde{\Psi}_I(\vb{R}_{i}\pm \delta \mathbf{R})}{\tilde{\Psi}_I(\vb{R}_i)} e^{-[V_{10}(\mathbf{R}_{i}\pm \delta \mathbf{R}) + V_{01}(\mathbf{R}_{i}\pm\delta \mathbf{R})+V_{10}(\mathbf{R}_i) + V_{01}(\mathbf{R}_i]\delta\tau/4} e^{E_T\delta\tau}\,.
\end{align}
One of the two walkers is kept in the propagation according to a heat-bath sampling among the two normalized weights $w_\pm/(\sum_{\pm} w_\pm)$ and the average weight $\sum_{\pm} w_\pm/2$ is associated to the propagated configuration.

In terms of the scalar Green's function, the propagation of Equation~(\ref{eq:prop}) reads
\begin{align}
\Psi(\tau + \delta\tau, \mathbf{R}_{i+1} ) = \int d \mathbf{R}_i \left[\frac{G_{\delta\tau}(\vb{R}_{i+1},\vb{R}_{i})}{\tilde{G}_{\delta\tau}^I(\vb{R}_{i+1},\vb{R}_{i})}\right] \tilde{G}_{\delta\tau}^I(\vb{R}_{i+1},\vb{R}_{i})\Psi(\tau, \mathbf{R}_i )\, .
\label{eq:prop_scal}
\end{align}
Since the new positions are sampled according to $\tilde{G}_{\delta\tau}^I(\vb{R}_{i+1},\vb{R}_{i})$, we can conveniently define 
\begin{align}
\Psi(\tau + \delta\tau, \mathbf{R}_{i+1} ) = \frac{G_{\delta\tau}(\vb{R}_{i+1},\vb{R}_{i})}{\tilde{G}_{\delta\tau}^I(\vb{R}_{i+1},\vb{R}_{i})} \Psi(\tau, \mathbf{R}_i )\,.
\label{eq:walker_I}
\end{align}
The imaginary-time evolution of the walker density is given by 
\begin{align}
& \Psi_I^\dagger(\mathbf{R}_{i+1}) \Psi(\tau + \delta\tau, \mathbf{R}_{i+1} ) = \nonumber\\
& \qquad \int d\mathbf{R}_i \left[ \frac{\Psi_I^\dagger(\mathbf{R}_{i+1}) G_{\delta\tau}(\vb{R}_{i+1},\vb{R}_{i})  \Psi(\tau, \mathbf{R}_i)}{\Psi_I^\dagger(\mathbf{R}_{i}) \tilde{G}_{\delta\tau}^I(\vb{R}_{i+1},\vb{R}_{i})\Psi(\tau, \mathbf{R}_i)}\right] \tilde{G}_{\delta\tau}^I(\vb{R}_{i+1},\vb{R}_{i})\Psi_I^\dagger(\mathbf{R}_{i}) \Psi(\tau, \mathbf{R}_i )\,.
\label{eq:prop_I_2}
\end{align}
Iterations of Equation~(\ref{eq:prop_I_2}) amount to multiple matrix multiplications
\begin{align}
\Psi(\tau,\mathbf{R}_N)=\left[\frac{G_{\delta\tau}(\vb{R}_{N},\vb{R}_{N-1})}{\tilde{G}_{\delta\tau}^I(\vb{R}_N,\vb{R}_{N-1})}\right] 
\left[\frac{G_{\delta\tau}(\vb{R}_{N-1},\vb{R}_{N-2})}{\tilde{G}_{\delta\tau}^I(\vb{R}_{N-1},\vb{R}_{N-2})}\right]\cdots 
\left[\frac{G_{\delta\tau}(\vb{R}_1,\vb{R}_0)}{\tilde{G}_{\delta\tau}^I(\vb{R}_1,\vb{R}_0)}\right] \Psi_T(\mathbf{R}_0)\,,
\end{align}
that are performed using the same matrices used to construct $|\Psi_T\rangle$. It has to be stressed that $\Psi(\tau,\mathbf{R}_N)$ is {\it not} the ground-state wave function. It rather represents a spin-isospin set of amplitudes that, when taken in product with the Hermitian conjugate of the importance function, gives an overlap for each component of the wave function. Are the changes in these overlaps that drive the distribution of walkers toward that of the true ground state. 

To avoid sign fluctuations in $\Psi_I^\dagger(\mathbf{R}_{i}) \Psi(\tau, \mathbf{R}_{i} )$, we sample the walkers from the positive-definite density distribution
\begin{align}
I(\mathbf{R}_i)= \left| \sum_{i_s,i_t} \langle \Psi_I(\mathbf{R}_{i}) |\chi_{i_s}\, \chi_{i_t} \rangle \langle \chi_{i_s}\, \chi_{i_t}| \Psi(\tau, \mathbf{R}_{i} ) \rangle \right| 
+\epsilon \sum_{i_s,i_t} \Big| \langle \Psi_I(\mathbf{R}_{i}) |\chi_{i_s}\, \chi_{i_t} \rangle \langle \chi_{i_s}\, \chi_{i_t} | \Psi(\tau, \mathbf{R}_{i} ) \rangle \Big|\,.
\label{eq:postive_w}
\end{align}
The first term simply measures the magnitude of the overlap of the wave functions, while the second, with a small coefficient $\epsilon\simeq 0.01$, ensures a positive
definite importance function to allow diffusion across nodal surfaces. This choice for the sampling distribution is monitored by checking how much this estimate of the population size deviates from the actual number of configurations. As alluded to earlier, similarly to standard Fermion diffusion Monte Carlo algorithms, the GFMC method suffers from the Fermion sign problem. Since the configurations are distributed according to $I(\mathbf{R}_i)$ defined in Equation~(\ref{eq:postive_w}),  the 
expectation values of observables that commute with the Hamiltonian are estimated as
\begin{align}
\langle \mathcal O(\tau) \rangle =\frac{\langle \Psi_T| \mathcal O |\Psi(\tau)\rangle}{\langle \Psi_T| \Psi(\tau)\rangle}=
 \frac{\sum_{R_i} \langle \Psi_T(\mathbf{R}_i)| O |\Psi(\tau,\mathbf{R}_i)\rangle / I(\mathbf{R}_i)}{\sum_{R_i} \langle \Psi_T(\mathbf{R}_i)| \Psi(\tau,\mathbf{R}_i)\rangle / I(\mathbf{R}_i)}\,.
\label{eq:O_exp}
\end{align}
For all other observables, we compute the mixed estimates
\begin{align}
\langle \mathcal O(\tau) \rangle \simeq 2 \frac{\langle \Psi_T | \mathcal O | \Psi(\tau)\rangle}{\langle \Psi_T | \Psi(\tau)\rangle} - \frac{\langle \Psi_T | \mathcal O | \Psi_T\rangle}{\langle \Psi_T | \Psi_T\rangle}\, ,
\label{eq:O_exp_nc}
\end{align}
where the first and the second term correspond to the DMC and VMC expectation value, respectively. 

As in standard Fermion diffusion Monte Carlo algorithms, the GFMC method suffers from the Fermion sign problem that arises from stochastically evaluating the matrix elements in Equation~(\ref{eq:O_exp}). The imaginary-time propagator is a local operator, while antisymmetry is a global property of the system. As a consequence, $|\Psi(\tau)\rangle$ can contain bosonic components that have much lower energy than the Fermionic ones and are exponentially amplified during the propagation. When the dot product with the antisymmetric $\Psi_T$ is taken, the desired Fermionic component is projected out in the expectation values, but the variance -- and hence the statistical error -- grows exponentially with $\tau$. Because the number of pairs that can be exchanged grows with $A$, the sign problem also grows exponentially with the number of nucleons. Already for $A=8$, the statistical errors grow so fast that convergence cannot be achieved.

To control the sign problem, we adopt the so-called ``contrained-path'' method~\cite{Wiringa:2000gb}, originally developed to study condensed matter systems~\cite{Zhang:1995zz}. This method is based on discarding those configurations that in future generations will contribute only noise to expectations values. If we knew the exact ground state, we could discard any walker for which
\begin{align}
\Psi_0^\dagger(\mathbf{R}_i) \Psi(\tau, \mathbf{R}_i ) = 0\,,
\end{align}
where a sum over spin-isospin states is implied. The sum of these discarded configurations can be written as a state $|\Psi_d\rangle$, which has zero overlap with the ground state. Disregarding $|\Psi_d\rangle$ is justified because it only contains excited-states components and should decay away as $\tau\to\infty$. However, in general, the exact ground state is not known, and the constraint is approximately imposed using $\Psi_T$ in place of $\Psi_0$:
\begin{align}
\langle \Psi_T| \Psi_d\rangle =0\, .
\label{eq:overlap}
\end{align}
The GFMC wave function evolves smoothly in imaginary time and changes can be made arbitrarily small by reducing $\delta\tau$. Hence, if the wave function is purely scalar, any configuration which yields a negative overlap must first pass through a point at which $\Psi_T$ -- and hence the overlap -- is zero. Discarding these configurations is then sufficient to stabilize the simulation and produce ``fixed-node'' variational solutions, to the many-Fermion problem. However, the GFMC trial wave function is a vector in spin-isospin space, and there are no coordinates for which all the spin-isospin amplitudes vanish. In addition, the overlap $\Psi_{T,\,p}^\dagger(\mathbf{R}_i) \Psi(\tau, \mathbf{R}_i )$ is complex and depends on the particular sampled order $p$. As a consequence, it does not evolve smoothly and can pass through zero. The constraint of Equation~(\ref{eq:overlap}) cannot be satisfied for individual configurations, but rather only on average for the sum of discarded configurations. To circumvent these difficulties, we define the overlap
\begin{align}
O_{T,\,p}= \Re[\Psi_{T,\,p}^\dagger(\mathbf{R}_i) \Psi(\tau, \mathbf{R}_i ) ]\,.
\end{align}
We can then introduce a probability for discarding a configuration in terms of the ratio $O_{T,\,p}/I_{T,\,p}$ where $I_{T,\,p}$ corresponds to choosing the ordering $p$ in $\Psi_I$ as defined in Equation~(\ref{eq:postive_w})
\begin{align}
P[\Psi_{T,\,p}^\dagger(\mathbf{R}_i), \Psi(\tau, \mathbf{R}_i )]=
\left\{
\begin{array}{cc}
0 & O/I > \alpha_c\nonumber\\
\frac{\alpha_C - O/I}{\alpha_c-\beta_c} & \alpha_c> O/I > \beta_c\nonumber\\
1 & O/I < \beta_c
\end{array} \right.
\end{align}
The constants $\alpha_c$ and $\beta_c$ are adjusted such that the average of the overlap $O_{T,\,p}/I_{T,\,P}$ is zero within statistical errors. 

In a few cases the constrained propagation converges to the wrong energy (either above or below the correct energy). Therefore, a small number, $n_u = 10$ to $80$, of unconstrained steps are made before evaluating expectation values. These few unconstrained steps appear to be sufficient to remove the bias introduced by the constraint but do not greatly increase the statistical error.

\subsection{Auxiliary field diffusion Monte Carlo}
Over the last two decades, the auxiliary field diffusion Monte Carlo method~\cite{Schmidt:1999lik} has become a mainstay for studying atomic nuclei~\cite{Lonardoni:2018prl,Lonardoni:2018prc,Lonardoni:2018nofk,Lynn:2020} and infinite neutron matter~\cite{Tews:2015ufa,Lynn:2015jua,Tews:2018kmu}. The AFDMC overcomes the exponential scaling with the number of nucleons of the GFMC by using a spin-isospin basis given by the outer product of single-nucleon spinors
\begin{align}
| \chi_{i_s}\, \chi_{i_t} \rangle \to |S\rangle \equiv  |s_1\rangle \otimes |s_2\rangle \otimes \dots \otimes |s_A\rangle \,,
\end{align}
where 
\begin{align}
|s_i\rangle = a_{i,\uparrow p} |\uparrow p\rangle + a_{i,\downarrow p} |\downarrow p\rangle + a_{i,\uparrow n} |\uparrow n\rangle + a_{i,\downarrow n} |\downarrow n\rangle\, .
\end{align}
The state vector is fully specified by a set of $4A$ complex coefficients. As opposed to the many-body spin-isospin basis defined in Equation~(\ref{eq:spin_isospin}), the single-particle one is not closed under the action of two-body operators. To see this, lets apply again the operator ${\bm \sigma}_1 \cdot {\bm \sigma}_2$ on a three-body spin state
\begin{align}
& {\bm \sigma}_1 \cdot {\bm \sigma}_2\Big[ \left( a_{1,\uparrow} |\uparrow \rangle +a_{1,\downarrow} |\downarrow \rangle  \right) \otimes 
\left( a_{2,\uparrow} |\uparrow \rangle +a_{2,\downarrow} |\downarrow \rangle  \right) \otimes
\left( a_{3,\uparrow} |\uparrow \rangle +a_{3,\downarrow} |\downarrow \rangle  \right) \Big] = \nonumber\\
& \qquad = 2 \Big[ \left( a_{2,\uparrow} |\uparrow \rangle +a_{2,\downarrow} |\downarrow \rangle  \right) \otimes 
\left( a_{1,\uparrow} |\uparrow \rangle +a_{1,\downarrow} |\downarrow \rangle  \right) \otimes
\left( a_{3,\uparrow} |\uparrow \rangle +a_{3,\downarrow} |\downarrow \rangle  \right) \Big] \nonumber\\
& \qquad - \Big[ \left( a_{1,\uparrow} |\uparrow \rangle +a_{1,\downarrow} |\downarrow \rangle  \right) \otimes 
\left( a_{2,\uparrow} |\uparrow \rangle +a_{2,\downarrow} |\downarrow \rangle  \right) \otimes
\left( a_{3,\uparrow} |\uparrow \rangle +a_{3,\downarrow} |\downarrow \rangle  \right) \Big] \,.
\end{align}
In general, the action of all pairwise spin/isospin operators needed to construct the trial state defined in Equation~(\ref{eq:psi_T}) generates all the $2^A\binom{A}{Z}$ amplitudes of the many-body spin-isospin basis. For this reason, the trial wave function typically used in AFDMC calculations~\cite{Gandolfi:2014ewa,Lonardoni:2018prc} is simpler than the one of the GFMC and takes the form
\begin{align}
|\Psi_T \rangle = \Big(1- \sum_{i<j} F_{ij} - \sum_{i<j<k} F_{ijk}  \Big) | \Phi_J\rangle\, ,
\end{align}
where $F_{ij}$ and $F_{ijk}$ are defined in Equations~(\ref{eq:Fij}) and (\ref{eq:Fijk}), respectively. Since it contains a linearized version of spin/isospin-dependent two-body correlations, this wave function is significantly cheaper to evaluate than the one used in GFMC, as it scales polynomially with the number of nucleons rather than exponentially. However, because only pairs of nucleons are correlated at a time, the cluster property is violated. Nevertheless, the use of these linearized spin-dependent correlations has enabled a number of remarkably accurate AFDMC calculations, in which properties of atomic nuclei up to $A=16$~\cite{Lonardoni:2018prl,Lonardoni:2018prc,Lonardoni:2018nofk} have been investigated utilizing the local $\chi$EFT interactions of Refs.~\cite{Gezerlis:2013ipa,Lynn:2015jua}. Very recently, the AFDMC trial wave function has been improved by including quadratic pair correlations~\cite{Lonardoni:2018prc,Lonardoni:2019}. 

The Jastrow component of $|\Psi_T \rangle$ is also simpler than the one of Equation~(\ref{def:phiJ_gfmc}),
\begin{align}
| \Phi_J\rangle=\prod_{i<j} f_{ij}^c  \prod_{i<j<k} f^{c}_{ijk} |\Phi_A(J^\pi, J_z, T_z)\rangle\,,
\label{def:phiJ_afdmc}
\end{align}
where the two-body scalar correlation are obtained consistently with the $u^p(r_{ij})$ minimizing the two-body cluster energy. The three-body scalar correlation is the one defined in Equation~(\ref{eq:fc_ijk}). The mean-field component is modeled by a sum of Slater determinants,
\begin{align}
\langle X| \Phi(J^\pi,J_z,T_z)\rangle=\sum_n c_n \left[\sum_{JJ_z} C_{JJ_z} \mathcal{A}\big[\,\phi_{\alpha_1}(x_1)\dots\phi_{\alpha_A}(x_A)\,\big] \right]_{JJ_z}\, .
\label{def:phi_afdmc}
\end{align}
In the above equation we have introduced $X=\{x_1,\dots,x_A\}$, where the generalized coordinate $x_i\equiv \{\mathbf{r}_i,s_i\}$ represents both the position $\mathbf{R} = \mathbf{r}_1,\dots,\mathbf{r}_A$ and the spin-isospin coordinates $S = s_1,\dots,s_A$ of the $A$ nucleons. The determinants are coupled with Clebsch-Gordan coefficients $C_{JJ_z}$ in order to reproduce the total angular momentum, total isospin, and parity. The single-particle orbitals are given by 
\begin{align} 
\phi_{\alpha}(x_i) = R_{nl}(r_i)\,Y_{ll_z}(\hat r_i)\,\chi_{ss_z}(\sigma)\,\chi_{tt_z}(\tau)\, ,
\label{eq:spo}
\end{align}
where $R_{nl}(r)$ is the radial function, $Y_{ll_z}$ is the spherical harmonic, and $\chi_{ss_z}(\sigma)$ and $\chi_{tt_z}(\tau)$ are the complex spinors describing the spin and isospin of the single-particle state. 

The AFDMC imaginary-time propagation can be broken up in small time steps similarly to what is done in Equation~(\ref{eq:path}) for the GFMC method. This time however, the generalized coordinate $X$ is used instead of $\mathbf{R}$ and the spin-isospin degrees of freedom are also sampled. The AFDMC wave function at imaginary time $\tau+\delta\tau$ can be written in an integral form analogous to the one of Equation~(\ref{eq:prop})
\begin{align}
\Psi(\tau + \delta\tau, X_{i+1} ) = \sum_{S_i} \int d \mathbf{R}_i G_{\delta\tau}(X_{i+1},X_i)\Psi(\tau, X_i )\, .
\label{eq:prop_afdmc}
\end{align}
Using the Trotter decomposition of Equation~(\ref{eq:trotter}), the short-time Green's function factorizes as 
\begin{align}
G_{\delta\tau}(X_{i+1},X_i) = G_{\delta\tau}^0(\vb{R}_{i+1},\vb{R}_{i}) \langle S_{i+1} | e^{-(V(\vb{R}_{i+1})/2 + V(\vb{R}_i)/2 -E_T)\delta\tau} |S_i\rangle +o(\delta\tau^3)\,.
\label{eq:prop_afdmc2}
\end{align}
Quadratic spin-isospin operators contained in the nuclear potential can connect a single spin-isospin state $|S_i\rangle$ to all possible $|S_{i+1}\rangle$ states. In order to preserve the single-particle representation, the short-time propagator is linearized utilizing the Hubbard-Stratonovich transformation
\begin{align}
e^{-\lambda \mathcal O^2 \delta\tau/2} = \frac{1}{\sqrt{2\pi}} \int_{-\infty}^\infty dx\, e^{-x^2/2}\, e^{x\sqrt{- \lambda \delta\tau}\, \mathcal O}\,,
\label{eq:V_prop}
\end{align}
where $x$ are the {\it auxiliary fields} and the operators $\mathcal O$ are obtained as follows. The first six terms defining the $N\!N$ potential can be conveniently separated in a spin/isospin-dependent $V_{SD}$ and spin/isospin-independent $V_{SI}$ contributions. To see this in more details, lets consider purely neutron systems, where $\bm\tau_i\cdot\bm\tau_j=1$, since the extension  to isospin-dependent terms is trivial~\cite{Lonardoni:2018prc}. In this case, $V_{SD}$ can be cast in the form 
\begin{align}
V_{SD}=\frac{1}{2} \sum_{i\alpha j \beta} A_{i\alpha, j\beta}\,\sigma_i^\alpha\,\sigma_j^\beta = \frac{1}{2}\sum_{n=1}^{3A} \mathcal O^2_{n}\,\lambda_n\, ,
\end{align}
where the operators $\mathcal O_{n}$ are defined as
\begin{align}
\mathcal O_{n}=\sum_{i,\alpha} \sigma_i^\alpha\,\psi^n_{i\alpha}\,.
\end{align}
In the above equations $\lambda_n$ and  $\psi^n_{i\alpha}$ are the eigenvalues and eigenvectors of the matrix $A$. Hence, applying the exponential of the spin-dependent terms of the $N\!N$ interaction amounts to rotating the spin-isospin states of nucleons
\begin{align}
e^{-V(\vb{R}_i)\delta\tau/2} |S_i\rangle = \prod_n \frac{1}{\sqrt{2\pi}} \int dx_n e^{-x_n^2/2} e^{x_n\sqrt{- \lambda \delta\tau}\, O_n} |S_i\rangle \,,
\label{eq:hubbard_prop}
\end{align}
and the imaginary-time propagation is performed by sampling the auxiliary fields $\bar{x}_n$ from the Gaussian probability distribution 
\begin{align}
|S_{i+1}\rangle = \prod_n e^{\bar{x}_n\sqrt{- \lambda \delta\tau}\, O_n} |S_i\rangle\, .
\end{align} 

The spin-orbit term of the $N\!N$ potential -- $p=7$ in Equation~(\ref{eq:NN_pot}) -- is implemented in the propagator as described in reference~\cite{Sarsa:2003zu}, and appropriate counter terms are included to remove the spurious contributions of order $\delta\tau$. Presently, the isospin-dependent spin-orbit term of the $N\!N$ potential, corresponding to $p=8$ in Equation~(\ref{eq:NN_pot}), cannot be properly treated within the AFDMC algorithm, as its counter term contains cubic spin-isospin operators, preventing the straightforward use of the Hubbard-Stratonovich transformation. 

Importance sampling techniques are also routinely implemented in the AFDMC method -- in both the spatial coordinates and spin-isospin configurations -- to drastically improve the efficiency of the algorithm. To this aim, the propagator of Equation~(\ref{eq:prop_afdmc2}) is modified as
\begin{align}
G^I_{\delta\tau}(X_{i+1},X_i) = G_{\delta\tau}(X_{i+1},X_i) \frac{\Psi_I(X_{i+1})}{\Psi_I(X_i)}\,,
\end{align}
and we typically take $\Psi_I(X)=\Psi_T(X)$. At each time step, each walker is propagated sampling a $3A$-dimensional vector to shift the spatial coordinates and a set of auxiliary fields $\mathcal{X}$ from Gaussian distributions. To remove the linear terms coming from the exponential of both Equations~(\ref{eq:prop_free}) and (\ref{eq:hubbard_prop}), in analogy to the GFMC method, we consider four weights, corresponding to separately flipping the sign of the spatial moves and spin-isospin rotations
\begin{align}
w_i=\frac{\Psi_I(\pm \mathbf{R}_{i+1}, S_{i+1}(\pm \mathcal{X}))}{\Psi_I(\mathbf{R}_i,S_i)} \,.
\label{eq:weight}
\end{align}
In the same spirit as the GFMC algorithm, only one of the four configurations is kept according to a heat-bath sampling among the four normalized weights $w_i/W$, with $W=\sum_{i=1}^4 w_i /4 $ being the cumulative weight. The latter is then rescaled by
\begin{align}
W \to W e^{-[V_{SI}(\mathbf{R}_i)/2+V_{SI}(\mathbf{R}_{i+1})/2-E_T]\delta\tau}\,,
\end{align}
and associated to this new configuration for branching and computing observables. This ``plus and minus'' procedure, first implemented in the AFDMC method in reference~\cite{Gandolfi:2014ewa} significantly reduces the dependence of the results on $\delta\tau$.

Expectation values are estimated during the imaginary-time propagation in a similar fashion as for the GFMC
\begin{align}
\langle \mathcal O(\tau) \rangle =\frac{\langle \Psi_T| \mathcal O |\Psi(\tau)\rangle}{\langle \Psi_T| \Psi(\tau)\rangle}=
 \frac{\sum_{X_i} \langle \Psi_T(X_i)| \mathcal O |\Psi(\tau,X_i)\rangle / \Psi_I(X_i)}{\sum_{X_i} \langle \Psi_T(X_i)| \Psi(\tau,X_i)\rangle / \Psi_I(X_i)}\,,
\label{eq:O_exp_afdmc}
\end{align}

\begin{figure}[tb]
\centering
\includegraphics[width=0.75\columnwidth]{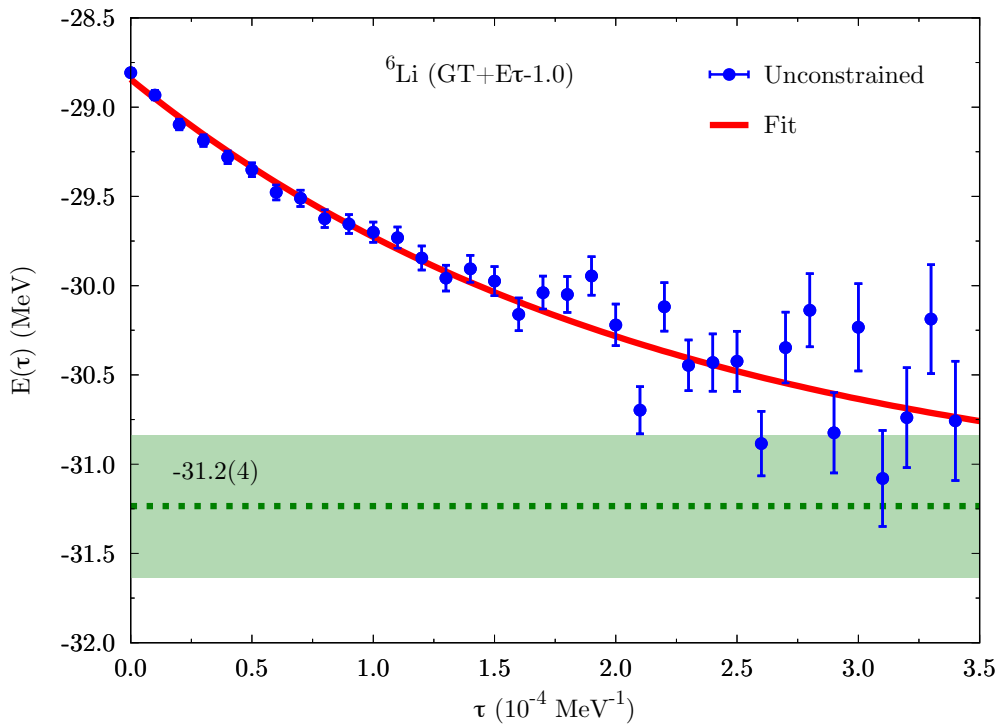}
\caption[]{Unconstrained evolution in $^6$Li for the GT+E$\tau$-1.0 local chiral Hamiltonian. Blue dots with error bars are AFDMC energies. The red curve is a single-exponential function fit to the AFDMC results. The green band represents the final energy result including the uncertainty coming from the fitting procedure.}
\label{fig:tr_li6}
\end{figure}

To alleviate the sign problem, as in reference~\cite{Zhang:2003zzk}, we implement an algorithm similar to the constrained-path approximation~\cite{Zhang:1996us}, but applicable to complex wave functions and propagators. The weights $w_i$ of Equation~(\ref{eq:weight}) are evaluated with
\begin{align}
\frac{\Psi_I(X_{i+1})}{\Psi_I(X_i)} \to {\rm Re} \left\{ \frac{\Psi_I(X_{i+1})}{\Psi_I(X_i)} \right\}\, ,
\end{align}
and they are set to zero if the ratio is negative. Unlike the fixed-node approximation, which is applicable for scalar potentials and for cases in which a real wave function can be used, the solution obtained from the constrained propagation is not a rigorous upper-bound to the true ground-state energy~\cite{Wiringa:2000gb}. To remove the bias associated with this procedure, the configurations obtained from a constrained propagation are further evolved using the following positive-definite importance sampling function~\cite{Pederiva:2004iz,Lonardoni:2018prc}
\begin{align}
\Psi_I(X) = \big|{ \rm Re} \{\Psi_T(X)\}\big| + \alpha\,\big|{ \rm Im} \{\Psi_T(X)\}\big|\,,
\end{align}
where we typically take $0.1 < \alpha < 0.5$. Along this unconstrained propagation, the expectation value of the energy is estimated according to Equation~(\ref{eq:O_exp_afdmc}). The asymptotic value is found by fitting the imaginary-time behavior of the unconstrained energy with a single-exponential function, as in reference~\cite{Pudliner:1997ck}. Unconstrained propagations have been performed in the latest AFDMC studies of atomic nuclei~\cite{Lonardoni:2018prl,Lonardoni:2018prc} and infinite nucleonic matter~\cite{Piarulli:2019pfq,Lonardoni:2019}. An example of unconstrained propagation in \isotope[6]{Li} for the GT+E$\tau$-1.0 local chiral Hamiltonian is reported in {\bf Figure~\ref{fig:tr_li6}}, where the blue dots with error bars are the AFDMC unconstrained energies, the red curve is the exponential fit, and the green band represents the final result including the uncertainty coming from the fitting procedure.

In summary, the VMC method is used to find the best possible guess for the wave function for a given nucleus, i.e., it is used to optimize the wave function variational parameters. VMC energies are usually above the ones coming from GFMC and AFDMC calculations, while other observables such as radii and density distributions are in closer agreement. The variationally optimized wave function is then used as input for the (statistically) exact GFMC and AFDMC algorithms. The difference between these two methods relies in their accuracy and limitations. The GFMC method is very accurate in predicting several observables with very small statistical error bars, but its applicability is limited up to 12 nucleons. The AFDMC method can tackle larger systems, but its precision is somewhat reduced and it is currently limited to somewhat simplified interactions~\cite{Carlson:2015}.

\section{Nuclear Structure Results} 
\begin{figure}[tb]
\centering
\includegraphics[width=\columnwidth]{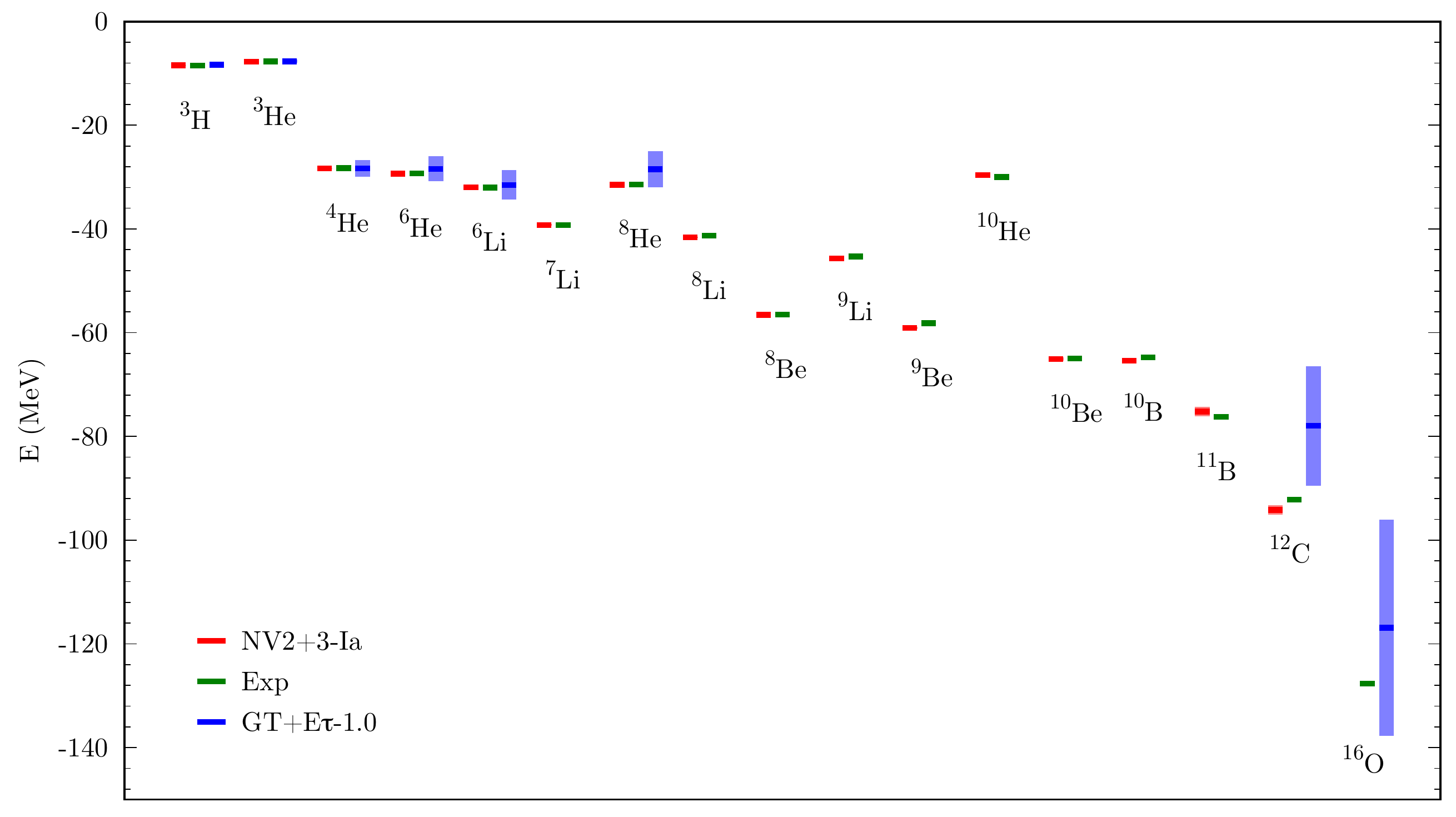}
\caption[]{Ground-state energies in $A\le16$ nuclei. For each nucleus, experimental results~\cite{Wang:2017} are shown in green at the center. GFMC (AFDMC) results for the NV2+3-Ia~\cite{Piarulli:2017dwd} (GT+E$\tau$-1.0~\cite{Lonardoni:2018prc}) potential are shown in red (blue) to the left (right) of the experimental values. For the NV2+3-Ia (GT+E$\tau$-1.0) potential, the colored bands include statistical (statistical plus systematic) uncertainties.}
\label{fig:ene}
\end{figure}

GFMC and AFDMC are complimentary methods that have been extensively used in the past to accurately calculate ground-state properties of light nuclei $(A\lesssim16)$. In the following we will present results obtained using the GFMC method for $\Delta$-full $\chi$EFT potentials, and using the AFDMC method for $\Delta$-less $\chi$EFT interactions. In {\bf Figure~\ref{fig:ene}} we show the binding energies of nuclei up to \isotope[16]{O} as calculated with GFMC for the NV2+3-Ia potential (red, left)~\cite{Piarulli:2017dwd}, and with AFDMC for the GT+E$\tau$-1.0 interaction (blue, right)~\cite{Lonardoni:2018prl,Lonardoni:2018prc}. The central green bars are the experimental data. GFMC results only carry Monte Carlo statistical uncertainties, while for AFDMC results, theoretical uncertainties coming from the truncation of the chiral expansion are also included. For an observable $X^{(i)}$ at order $i=0,2,3,\ldots$, the theoretical uncertainty $\delta X^{(i)}$ is estimated according to the prescription of Epelbaum \textit{et al.}~\cite{Epelbaum:2015}:
\begin{align}
    \delta X^{(0)}&=Q^2\big|X^{(0)}\big|\,, \nonumber \\[0.2cm]
    \delta X^{(i)}&=\max_{2\le j\le i}\left(Q^{i+1}\big|X^{(0)}\big|,\,Q^{i+1-j}\big|\Delta X^{(j)}\big|\right)\quad \text{for }i\ge 2\,, \nonumber \\[0.2cm]
    \delta X^{(i)}&\ge\max\left(\big|X^{(j\ge i)}-X^{(k\ge i)}\big|\right)\,, 
\end{align}
where
\begin{align}
    \Delta X^{(2)}&\equiv X^{(2)}-X^{(0)}, \nonumber \\
    \Delta X^{(i)}&\equiv X^{(i)}-X^{(i-1)}\quad \text{for }i\ge 3\,.
\end{align}
For the local chiral interaction GT+E$\tau$-1.0, results are presented at N$^2$LO $(i=3)$ considering $Q=m_\pi/\Lambda_b$, with $m_\pi\approx140\,\rm MeV$ and $\Lambda_b=600\,\rm MeV$~\cite{Lonardoni:2018prl,Lonardoni:2018prc}.

\begin{figure}[tb]
\centering
\includegraphics[width=\columnwidth]{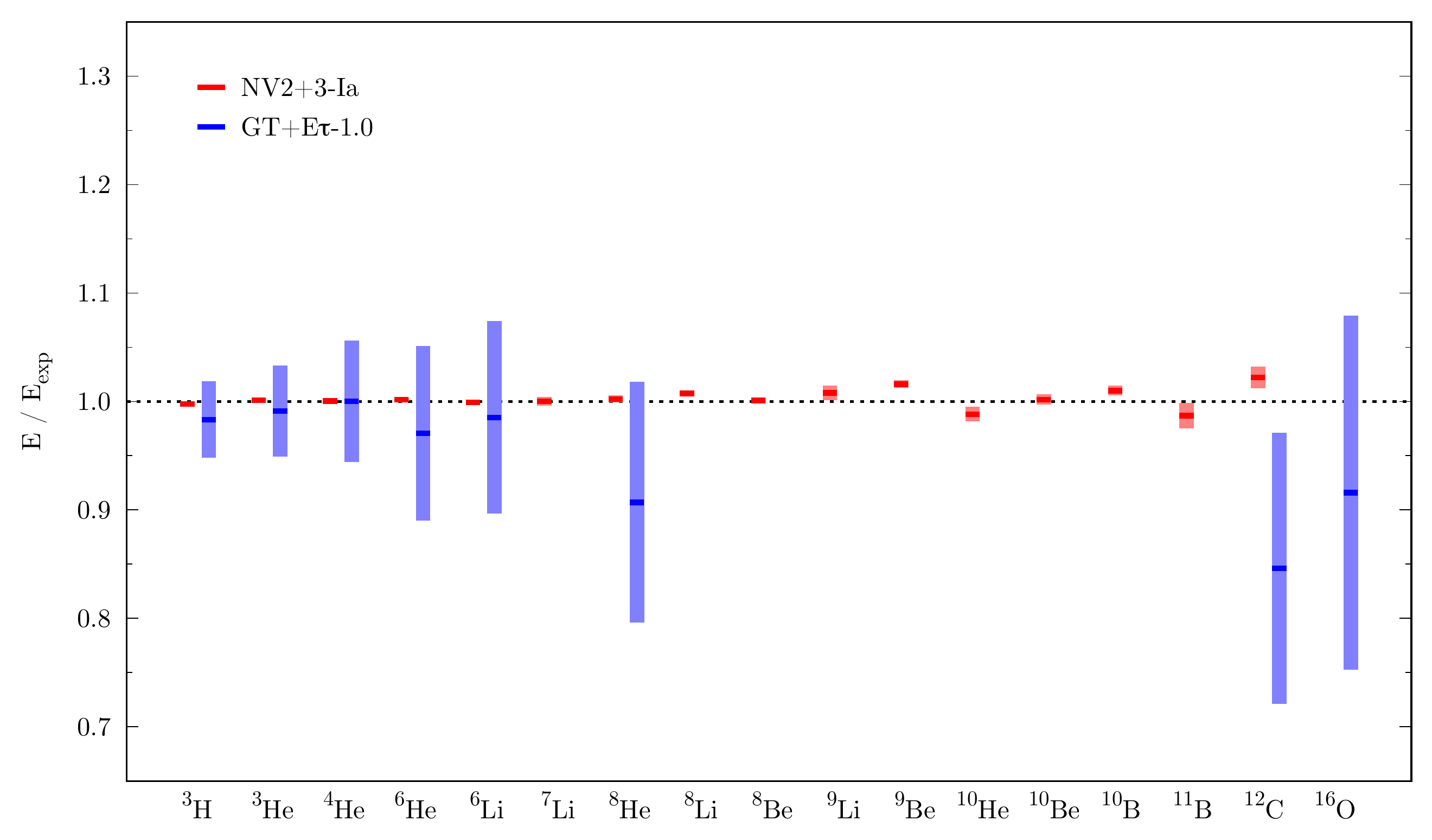}
\caption[]{Energy ratio between the calculated binding energies and the experimental data. The color scheme is the same as {\bf Figure~\ref{fig:ene}}.}
\label{fig:ene_rat} 
\end{figure}

The NV2+3-Ia interaction provides an overall good description of the
ground-state energy of light nuclei, including neutron-rich systems
with isospin asymmetry as large as 0.6 (\isotope[10]{He}). This can be
appreciated even more by looking at {\bf Figure~\ref{fig:ene_rat}}, where the ratio
between QMC results and experimental data is shown. Above $A=8$, the
NV2+3-Ia description of binding energies looks slightly less accurate,
with some nuclei slightly underbound (\isotope[10]{He, \isotope[11]{B}})
and some other sightly overbound (\isotope[9]{Be}, \isotope[10]{B},
\isotope[12]{C}). However, the difference with the experimental values
is always less than $0.2\,{\rm MeV}/A$, discrepancy that we expect to be
fully covered by the uncertainty coming from the truncation of the chiral
expansion (i.e., theoretical uncertainty from the interaction
model), currently not available for the NV2+3-Ia potential.

The binding energy of very light nuclei is also well reproduced by the
GT+E$\tau$-1.0 interaction, with \isotope[8]{He} slightly underbound
($0.37\,{\rm MeV}/A$ difference compared to the experimental value),
but compatible with observations within the estimated statistical
plus systematic uncertainties, see {\bf Figure~\ref{fig:ene_rat}}. Differently
from GFMC calculations, AFDMC results for $8\le A\le11$ open-shell
nuclei are currently not available. The ground-state energy of heavier
closed-shell systems, such as \isotope[12]{C} and \isotope[16]{O}, for the
GT+E$\tau$-1.0 potential is higher than the expected result. However,
the binding energy of \isotope[16]{O} is still compatible with the
experimental value within the fully uncertainty estimate. As discussed in
reference~\cite{Lonardoni:2018prc}, the discrepancy found for \isotope[12]{C}
is due to the somewhat too simplistic $A=12$ AFDMC wave function, that
only includes couplings in the $p$-shells, rather than a deficiency of
the interaction itself. It has to be noted that AFDMC results for the
GT+E$\tau$-1.0 interaction carry larger overall uncertainties compared
to GFMC results for the NV2+3-Ia potential. This is because the full
uncertainty evaluation includes both statistical and theoretical
errors. Both QMC methods imply statistical uncertainties of the order
of few percent. For the $\Delta$-less potential, the theoretical errors
coming from the truncation of the chiral expansion dominate compared
to the statistical errors. Considering the next order in the chiral
expansion should reduce theoretical uncertainties, and work is
currently being done in developing such potentials.

\begin{figure}[tb]
\centering
\includegraphics[width=\columnwidth]{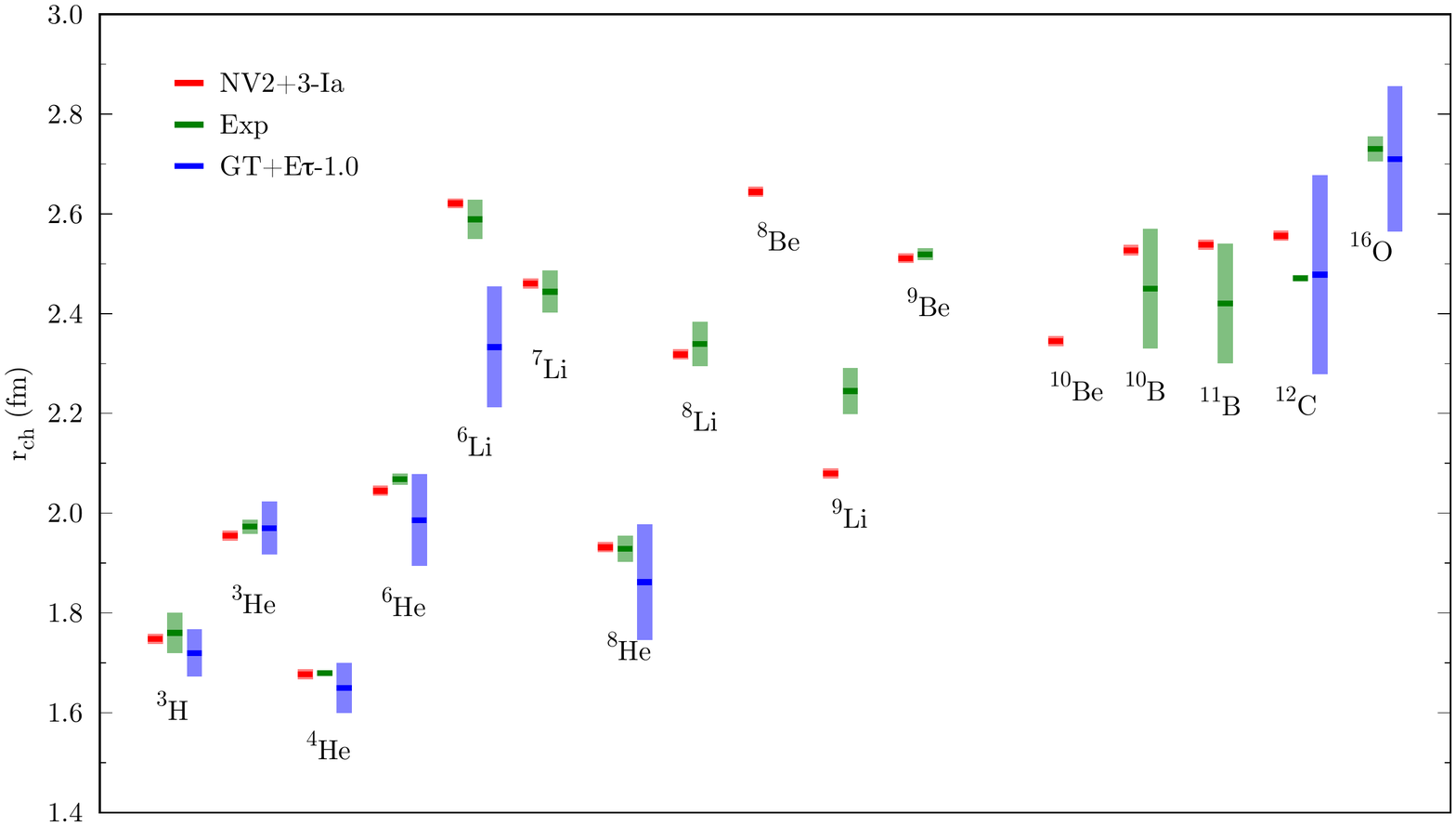}
\caption[]{Same as {\bf Figure~\ref{fig:ene}} but for charge radii. Experimental data are from references~\cite{Fricke:2004,Sick:2008,Mueller:2007,Nortershauser:2011}.}
\label{fig:rch}
\end{figure}

{\bf Figure~\ref{fig:rch}} shows the charge radii of $A\le16$ nuclei for the NV2+3-Ia and GT+E$\tau$-1.0 potentials, with respect to the available experimental data. The expectation value of the charge radius is derived from the point-proton radius $r_{\rm pt}$ using the relation
\begin{align}
	\left\langle r_{\rm ch}^2\right\rangle=
	\left\langle r_{\rm pt}^2\right\rangle+
	\left\langle R_p^2\right\rangle+
	\frac{A-Z}{Z}\left\langle R_n^2\right\rangle+
	\frac{3\hbar^2}{4M_p^2c^2}+
	\left\langle r_{\rm so}^2\right\rangle\,,
	\label{eq:rch}
\end{align}
where $\left\langle R_p^2\right\rangle=0.770(9)\,\rm{fm}^2$ is the proton radius~\cite{Beringer:2012}, $\left\langle R_n^2\right\rangle=-0.116(2)\,\rm{fm}^2$ is the neutron radius~\cite{Beringer:2012}, $(3\hbar^2)/(4M_p^2c^2)\approx0.033\,\rm{fm}^2$ is the Darwin-Foldy correction~\cite{Friar:1997}, and $\left\langle r_{\rm so}^2\right\rangle$ is a spin-orbit correction due to the anomalous magnetic
moment in halo nuclei~\cite{Ong:2010}. The point-nucleon radius $r_{\rm pt}$ is calculated as
\begin{align}
	\left\langle r_N^2\right\rangle=\frac{1}{{\mathcal N}}\big\langle\Psi\big|\sum_i\mathcal P_{N_i} |\vb{r}_i|^2\big|\Psi\big\rangle\,,
\end{align}
where $\vb{r}_i$ is the intrinsic coordinate of \cref{eq:r_intr}, ${\mathcal N}$ is the number of protons or neutrons, and 
\begin{align}
	\mathcal P_{N_i}=\frac{1\pm\tau_{z_i}}{2}
	\label{eq:proj}
\end{align}
is the projector operator onto protons ($+$) or neutrons ($-$). The charge radius is a mixed expectation value, and it requires the calculation of both VMC and DMC point-proton radii, according to Equation~(\ref{eq:O_exp_nc}). Even though mixed expectation values typically depend on the quality of the employed trial wave functions, for the highly-accurate wave functions employed in the GFMC and AFDMC methods, the extrapolation of the mixed estimate $\left\langle r_{\rm ch}^2\right\rangle$ is always small.

Both chiral interactions nicely reproduce the charge radius of helium
isotopes. The NV2+3-Ia potential also reproduces the radius of lithium,
beryllium, and boron isotopes, with new predictions for \isotope[8]{Be}
and \isotope[10]{Be}. The charge radius of \isotope[9]{Li} is
underpredicted, whereas that of \isotope[12]{C} is overestimated. The
GT+E$\tau$-1.0 potential works remarkably well in predicting the
charge radius of \isotope[12]{C} and \isotope[16]{O}, even though
theoretical uncertainties, that dominate over the statistical one,
are large. As discussed in the previous paragraphs,
going to the next order in the chiral expansion will reduce such
theoretical uncertainties. For the GT+E$\tau$-1.0 interaction, 
the charge radius of \isotope[6]{Li} turns out
to be smaller compared to the experimental value. Once again, this is not
a feature of the employed interaction, rather a deficiency of the AFDMC
wave function. In fact, differently from GFMC, the current AFDMC wave
function does not include dedicated $\alpha$-deuteron-like correlations,
necessary to capture the structural properties of \isotope[6]{Li}.

\begin{figure}[tb]
\centering
\includegraphics[width=0.75\columnwidth]{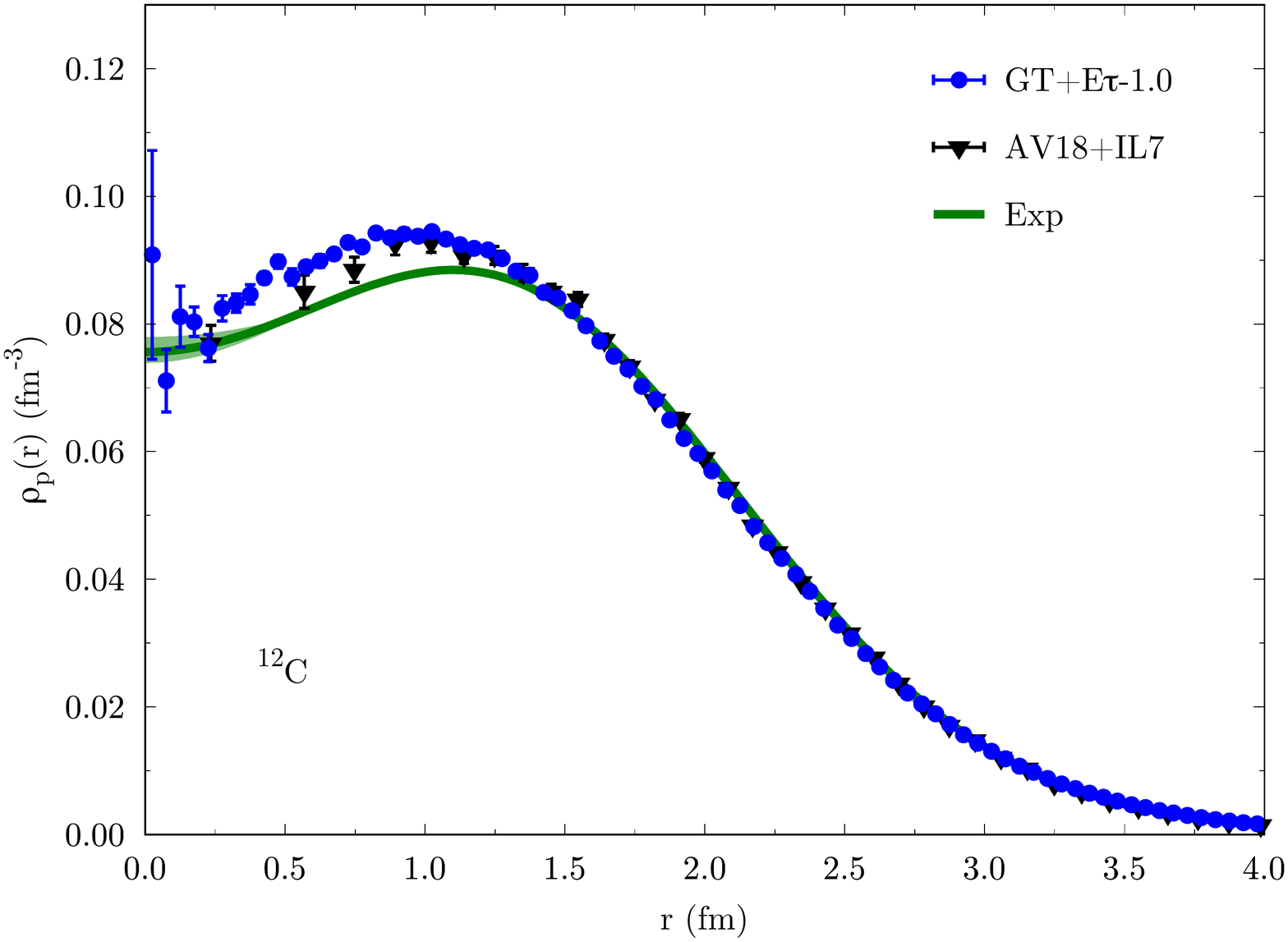}
\caption[]{Proton density in \isotope[12]{C}. Black triangles are GFMC results for the AV18+IL7 potential~\cite{Lovato:2013}. Blue dots are AFDMC results for the GT+E$\tau$-1.0 interaction~\cite{Lonardoni:2018prc}. The green band corresponds to the experimental results, unfolded from electron scattering data (see text for details).}
\label{fig:rho_c12}
\end{figure}

\begin{figure}[tb]
\centering
\includegraphics[width=0.75\columnwidth]{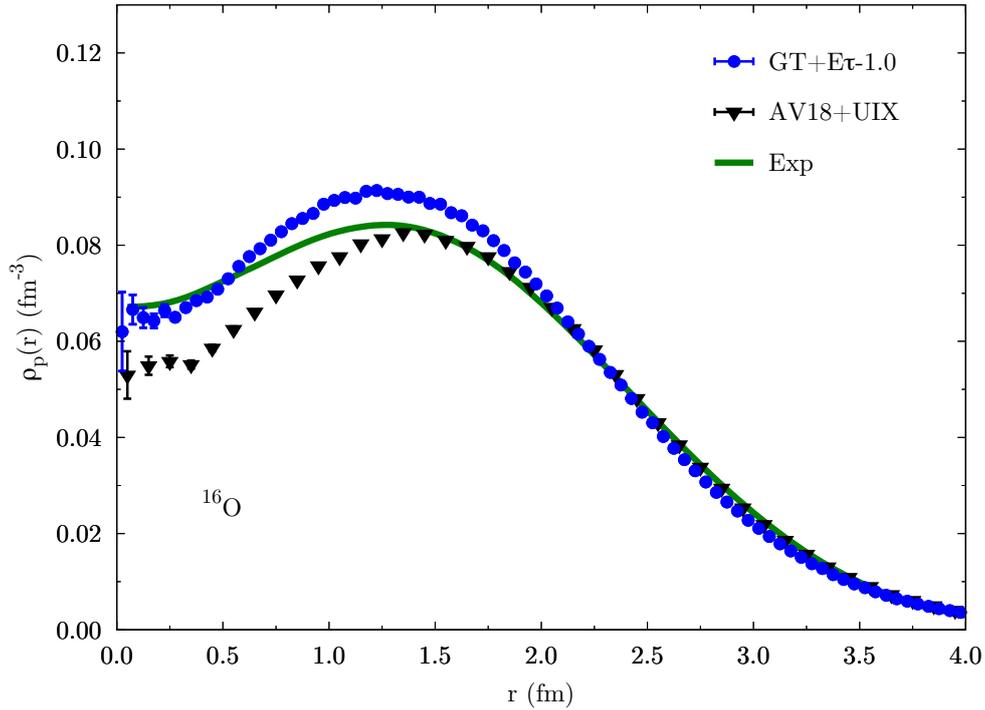}
\caption[]{Same as \cref{fig:rho_c12} but for \isotope[16]{O}. Black triangles are cluster VMC results for the AV18+UIX potential~\cite{Lonardoni:2017}. Blue dots are AFDMC results for the GT+E$\tau$-1.0 interaction~\cite{Lonardoni:2018prc}.}
\label{fig:rho_o16}
\end{figure}

\begin{figure}[tb]
\centering
\includegraphics[width=0.75\columnwidth]{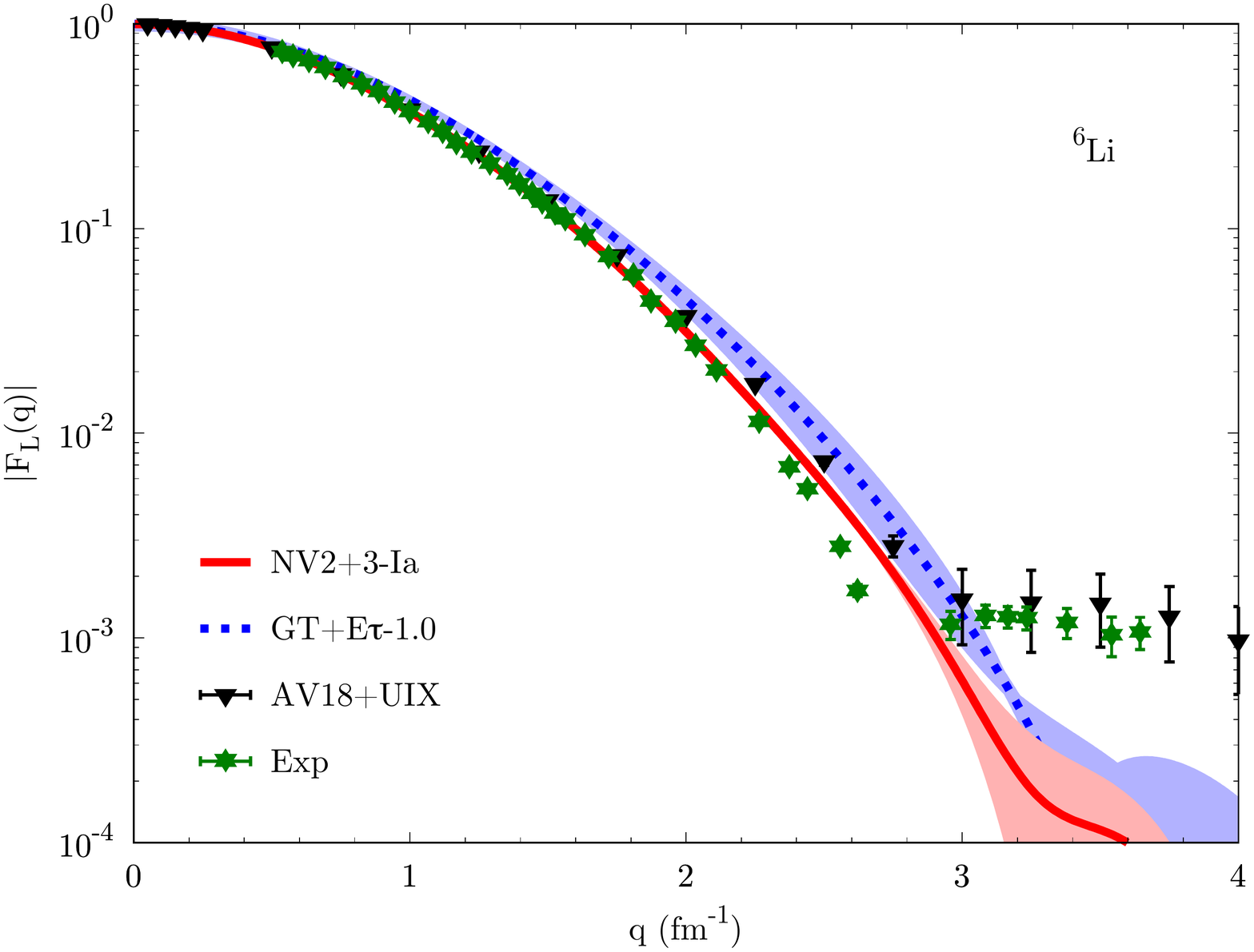}
\caption[]{Longitudinal elastic form factor in $^6$Li for different nuclear potentials. For the NV2+3-Ia (solid red line) and AV18+UIX (black triangles) potentials, errors correspond to statistical Monte Carlo uncertainties. The blue band for the GT+E$\tau$-1.0 potential also includes the uncertainties coming from the truncation of the chiral expansion. Green stars are the experimental values~\cite{Li:1971}. Adapted from reference~\cite{Lonardoni:2018prc}.}
\label{fig:ff_li6}
\end{figure}

In QMC methods, single-nucleon densities are calculated as
\begin{align}
	\rho_{N}(r) &=\frac{1}{4\pi r^2}\big\langle\Psi\big|\sum_i    \mathcal P_{N_i}\delta(r-|\vb{r}_i|)\big|\Psi\big\rangle\,, 
	\label{eq:rho_N}
\end{align}
where $\mathcal P_{N_i}$ is the projector operator of Equation~(\ref{eq:proj})
and $\rho_N$ integrates to the number of nucleons. In {\bf Figures \ref{fig:rho_c12}, \ref{fig:rho_o16}}
we show the QMC proton density in
\isotope[12]{C} and \isotope[16]{O} for the available phenomenological
(black) and chiral EFT (blue) potentials. Error bars correspond to
statistical uncertainties only. The green bands are the experimental
single-nucleon densities, obtained from the ``sum-of-Gaussians''
parametrization of the charge densities given in reference~\cite{Devries:1987}
by unfolding the nucleon form factors and subtracting the small
contribution of the neutrons. As can be seen, both phenomenological
and chiral EFT interactions provide a good description of the proton
density in \isotope[12]{C}. The small discrepancy with the experimental
curve at short distance is due to two-body meson exchange currents,
not included in the proton density presented here. As shown in
reference~\cite{Lovato:2013}, such currents have little effect on the
single-nucleon density for $A\ge12$, slightly reducing its value at
small $r$. The phenomenological AV18+UIX potential underestimates the
proton density a short distance in \isotope[16]{O}. As indicated by
the cluster VMC analysis of reference~\cite{Lonardoni:2017}, the three-body
potential UIX introduces repulsion in the system, pushing nucleons
far away from the nucleus center of mass, and thus resulting in larger
radius and smaller central density. The \isotope[16]{O} AFDMC density
for the GT+E$\tau$-1.0 potential is instead in better agreement with the
experimental curve. 

As opposed to the charge radius, densities are not observables themselves. However, the single-nucleon density can be related to the longitudinal elastic (charge) form factor, physical quantity experimentally accessible via electron-nucleon scattering processes. In fact, the charge form factor can be expressed as the ground-state expectation value of the one-body charge operator~\cite{Mcvoy:1962}, which, ignoring small spin-orbit contributions in the one-body current, results in the following expression:
\begin{align}
	F_L(q)=\frac{1}{Z}\frac{G_E^p(Q_{\rm el}^2)\,\tilde{\rho}_p(q)+G_E^n(Q_{\rm el}^2)\,\tilde{\rho}_n(q)}{\sqrt{1+Q_{\rm el}^2/(4 m_N^2)}}\,,
	\label{eq:ff}
\end{align}
where $\tilde{\rho}_{N}(q)$ is the Fourier transform of the single-nucleon density defined in~\cref{eq:rho_N}, and $Q^2_{\rm el}=\vb{q}^2-\omega_{\rm el}^2$ is the four-momentum squared, with $\omega_{\rm el}=\sqrt{q^2+m_A^2}-m_A$ the energy transfer corresponding to the electron scattering elastic peak, $m_A$ being the mass of the target nucleus. $G_E^N(Q^2)$ are the nucleon electric form factors, for which we adopt Kelly's parametrization~\cite{Kelly:2004}.

\begin{figure}[tb]
\centering
\includegraphics[width=0.75\columnwidth]{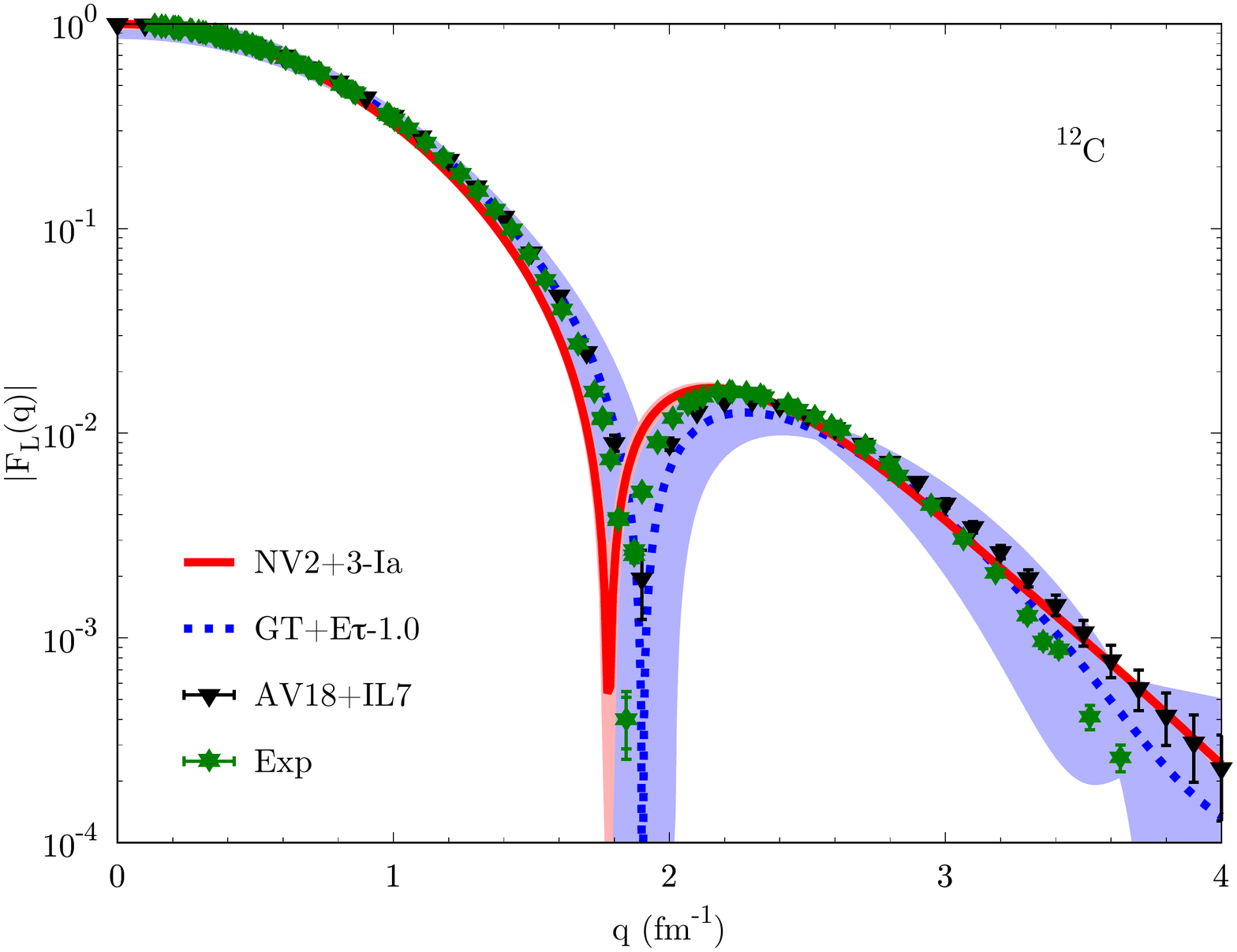}
\caption[]{Same as {\bf Figure~\ref{fig:ff_li6}} but for $^{12}$C. Experimental data are taken from reference~\cite{Devries:1987}. Adapted from reference~\cite{Lonardoni:2018prc}.}
\label{fig:ff_c12}
\end{figure}

\begin{figure}[tb]
\centering
\includegraphics[width=0.75\columnwidth]{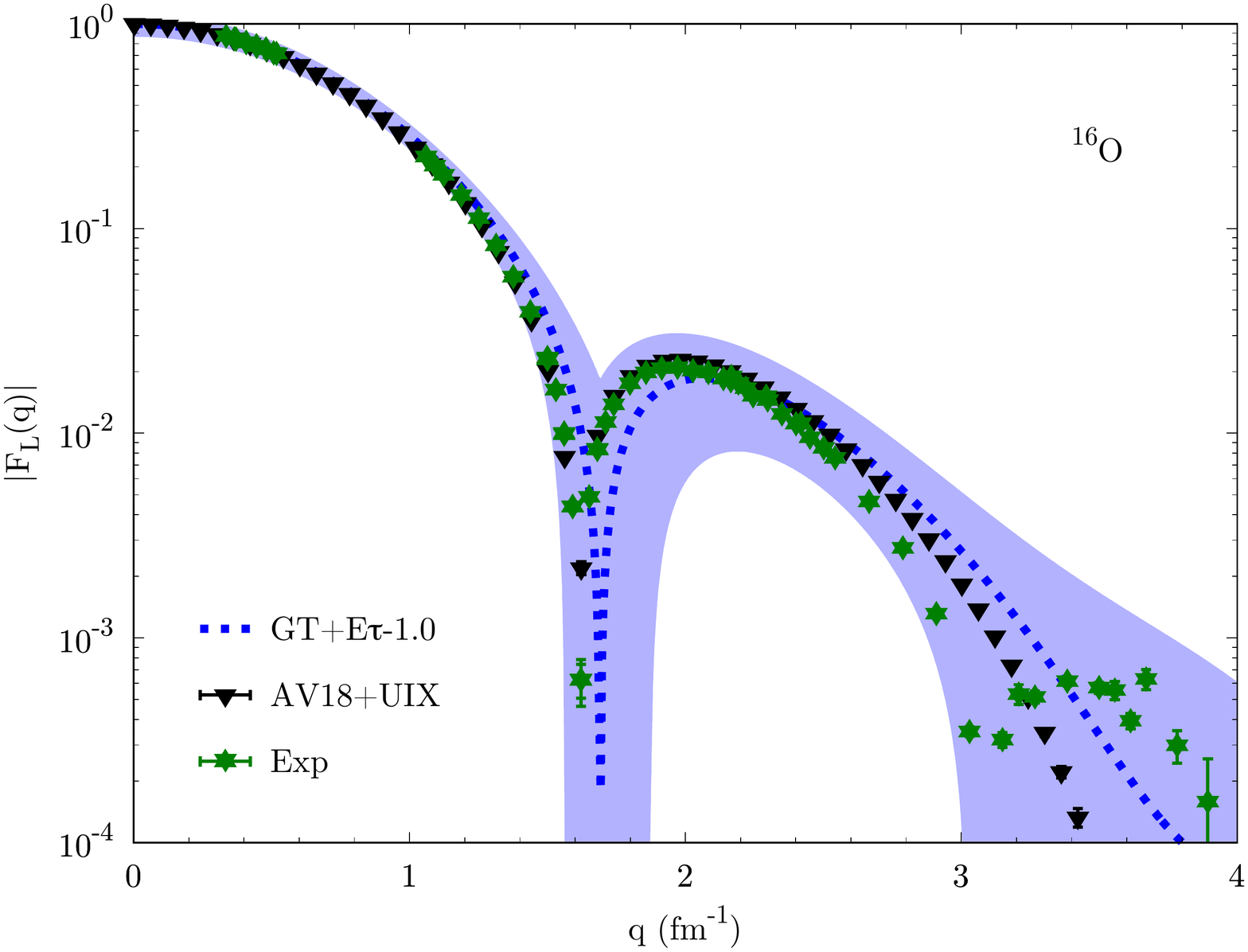}
\caption[]{Same as {\bf Figure~\ref{fig:ff_li6}} but for $^{16}$O. Experimental data are from Sick, based on references~\cite{Sick:1970,Schuetz:1975,Sick:1975}. Adapted from reference~\cite{Lonardoni:2018prc}.}
\label{fig:ff_o16}
\end{figure}

In {\bf Figures~\ref{fig:ff_li6}-\ref{fig:ff_o16}} we show the charge form
factor in \isotope[6]{Li}, \isotope[12]{C}, and \isotope[16]{O}. Lines
with bands correspond to chiral interactions, solid red for
NV2+3-Ia from GFMC calculations and dotted blue
for GT+E$\tau$-1.0 from AFDMC calculations. The
black triangles are the results for the phenomenological potentials:
AV18+UIX in \isotope[6]{Li} from VMC
calculations~\cite{Wiringa:1998}, AV18+IL7 in \isotope[12]{C} from
GFMC calculations~\cite{Lovato:2013}, and AV18+UIX in \isotope[16]{O}
from cluster VMC calculations~\cite{Lonardoni:2017}. Green
stars are the available experimental
results~\cite{Li:1971,Devries:1987,Sick:1970,Schuetz:1975,Sick:1975}. Note
that for all QMC calculations of the charge form factor only
one-body charge operators are considered, i.e., no
two-body electromagnetic currents are included. However, as shown
in references~\cite{Wiringa:1998,Lovato:2013,Mihaila:2000}, such operators
give a non-negligible contribution only for $q>2\,\rm fm^{-1}$, as they
basically include relativistic corrections.

In \isotope[6]{Li} all interactions provide a consistent description of the charge form factor, with NV2+3-Ia and AV18+UIX compatible with the experimental results up to $q\approx2\,\rm fm^{-1}$, where two-body currents start playing a role. In the same range, the GT+E$\tau$-1.0 results are slightly higher, as already indicated by the too small charge radius (see {\bf Figure~\ref{fig:rch}}). Interestingly, only the phenomenological potential is capable of reproducing the kink in the experimental data, while chiral interactions predict a smooth charge form factor also above $q\approx3\,\rm fm^{-1}$. The inclusion of two-body currents could improve the description of the charge form factor at high momentum. However, this is a momentum range roughly corresponding to the characteristic cut-off of chiral potentials, hence their description of observables in such regime is not supposed to hold. Similar conclusions can be drawn for the charge form factor in \isotope[12]{C} and \isotope[16]{O}, where chiral forces produce results compatible with the experimental data, in particular for the position of the first diffraction peak. This is slightly underestimated for \isotope[12]{C} with the NV2+3-Ia potential, but we expect it to be consistent with the experimental results once the uncertainties coming from the truncation of the chiral expansion are taken into consideration.

Note that the ``zero'' in the form factor is due to the presence of a term like $\sin^2(qR)$, where $R$ is related to the nucleus charge radius. The zero is obtained when $qR = \pi$. Therefore, a smaller (larger) $q$ value for the zero compared to the experimental data suggests a larger (smaller) $R$ value, i.e., a larger (smaller) $r_{\rm ch}$ value. This is indeed verified by QMC calculations. For instance, in {\bf Figure~\ref{fig:ff_c12}}, the NV2+3-Ia potential predicts a smaller $q$ value for the zero of the charge form factor in \isotope[12]{C}, hence a larger value for the charge radius, as confirmed by {\bf Figure~\ref{fig:rch}}.

\section*{Conclusions}
In this work we have reviewed recent advancements in the development of realistic nuclear interactions and of \textit{ab-initio} many-body methods for nuclear physics. In particular, we have discussed the recent integration of nearly-local interactions derived within chiral effective field theory, both with and without the inclusion of $\Delta$ degrees of freedom, in quantum Monte Carlo methods, namely variational Monte Carlo, Green's function Monte Carlo, and auxiliary field diffusion Monte Carlo. Such a successful combination lead to accurate and realistic calculations of ground- and excited-state properties of nuclei, that include but is not limited to spectra, charge radii, and longitudinal elastic form factors. Even though the chiral interactions discussed in this work have been constructed using few-body observables only, nucleon-nucleon scattering data and properties of nuclei up to $A=5$, they provide a remarkable description of the physics of nuclei up to, at least, $A=16$, with excellent agreement with experimental data. 

The same techniques and nuclear potentials reviewed here have also been used to calculate the equation of state of infinite nuclear and neutron matter~\cite{Piarulli:2019pfq,Lonardoni:2019}, and to infer properties of neutron stars, with results compatible with astrophysical observations including constraints extracted from gravitational waves of the neutron-star merger GW170817 by the LIGO-Virgo detection~\cite{Abbott:2018}.

Future efforts will be dedicated to i) further improve the employed local chiral interactions, by extending to higher order in the chiral expansion, ii) calculate electroweak properties in nuclear systems, by consistently deriving electroweak currents, and iii) extend the calculations to heavier nuclei, by improving the AFDMC variational wave functions and the scaling of the algorithm.

\section*{Author Contributions}
All authors listed have made a substantial, direct and intellectual contribution to the work, and approved it for publication.

\section*{Funding}
The work of S.G. was supported by the U.S. Department of Energy, 
Office of Science, Office of Nuclear Physics, under contract No.~DE-AC52-06NA25396, 
by the NUCLEI SciDAC program, by the LDRD program at LANL, and 
by the DOE Early Career research Program.
The work of D.L. was supported by the U.S. Department of Energy, 
Office of Science, Office of Nuclear Physics, under Contract No. DE-SC0013617, 
and by the NUCLEI SciDAC program.
The work of A.L. was supported by the U.S. Department of
Energy, Office of Science, Office of Nuclear Physics, under
contract No. DE-AC02-06CH11357. 
The work of M.P. was supported by the U.S. Department of Energy, 
Office of Science, Office of Nuclear Physics, under the FRIB Theory Alliance award DE-SC0013617.
Computational resources have been provided by the Los Alamos National Laboratory
Institutional Computing Program, which is supported by the U.S. Department
of Energy National Nuclear Security Administration under Contract
No.~89233218CNA000001, and by the National Energy Research Scientific
Computing Center (NERSC), which is supported by the U.S. Department of
Energy, Office of Science, under contract No.~DE-AC02-05CH11231.


\end{document}